\documentclass[9pt, a4paper]{article}

\usepackage{a4wide}
\usepackage{xcolor}
\usepackage{amssymb}
\usepackage{amsmath}
\usepackage{graphicx}
\usepackage{url}
\usepackage{rotating}
\usepackage{bm}
\usepackage[utf8]{inputenc}
\usepackage[parfill]{parskip}
\usepackage{lipsum}
\usepackage{lineno}
\usepackage{etoolbox}

\newcommand\blfootnote[1]{%
	\begingroup
	\renewcommand\thefootnote{}\footnote{#1}%
	\addtocounter{footnote}{-1}%
	\endgroup
}

\def\code#1{\texttt{#1}}

\newcommand{\beginsupplement}{%
	\setcounter{table}{0}
	\renewcommand{\thetable}{S.\arabic{table}}%
	\setcounter{figure}{0}
	\renewcommand{\thefigure}{S.\arabic{figure}}%
	\setcounter{section}{0}
	\renewcommand*{\thesection}{S.\the\value{section}}
	\setcounter{page}{1} 
	\resetlinenumber 
}

\usepackage{titling} 
\usepackage{authblk} 

\title{Multiscale Influenza Forecasting}

\author[ ]{Dave Osthus$^{*\text{a}}$}
\author[ ]{Kelly R. Moran$^{\text{a,b}}$}

\affil[a]{Los Alamos National Laboratory, Statistical Sciences Group}
\affil[b]{Duke University, Department of Statistical Science}

\date{}

\begin{document}

\maketitle

\begin{abstract}
\blfootnote{$^*$Email correspondence can be directed to dosthus@lanl.gov}Influenza forecasting in the United States (US) is complex and challenging for reasons including substantial spatial and temporal variability, nested geographic scales of forecast interest, and heterogeneous surveillance participation. 
Here we present a flexible influenza forecasting model called Dante, a multiscale flu forecasting model that learns rather than prescribes spatial, temporal, and surveillance data structure. 
Forecasts at the Health and Human Services (HHS) regional and national scales are generated as linear combinations of state forecasts with weights proportional to US Census population estimates, resulting in coherent forecasts across nested geographic scales. 
We retrospectively compare Dante's short-term and seasonal forecasts at the state, regional, and national scales for the 2012 through 2017 flu seasons in the US to the Dynamic Bayesian Model (DBM), a leading flu forecasting model. 
Dante outperformed DBM for nearly all spatial units, flu seasons, geographic scales, and forecasting targets. 
The improved performance is due to Dante making forecasts, especially short-term forecasts, more confidently and accurately than DBM, suggesting Dante's improved forecast scores will also translate to more useful forecasts for the public health sector. 
Dante participated in the prospective 2018/19 FluSight challenge hosted by the Centers for Disease Control and Prevention and placed 1st in both the national and regional competition and the state competition.
The methodology underpinning Dante can be used in other disease forecasting contexts where nested geographic scales of interest exist.
\end{abstract}

\section*{Introduction}
Influenza represents a significant burden to public health with an estimated 9 to 49 million cases each year in the United States (US) \cite{diseaseburden}.
Influenza (flu) related activity is monitored in the US by the Centers for Disease Control and Prevention (CDC) through numerous surveillance efforts. 
One such effort is the Outpatient Influenza-like Illness Surveillance Network (ILINet).
ILINet collects weekly data on influenza-like illness (ILI) from over 2000 healthcare providers from all 50 states, Puerto Rico, the US Virgin Islands, and the District of Columbia. 
ILI is defined as a temperature greater or equal to 100 degrees Fahrenheit, a cough or sore throat, and no other known cause, representing symptoms consistent with influenza. 
ILINet constitutes a significant and necessary effort to understanding the spread and prevalence of flu-like illness in the US in near real-time.

With mature ILI surveillance infrastructure in place in the US, attention has turned in recent years to ILI prediction. 
The ability to predict the spread of ILI poses a substantial public health opportunity if able to be done accurately, confidently, and with actionable lead times at geographic and temporal scales amenable to public health responsiveness. 
Since 2013, the CDC has hosted an influenza forecasting challenge called the FluSight challenge to gauge the feasibility of forecasting targets of public health interest in real-time, to galvanize the flu forecasting community around common goals, and to foster innovation and improvement through collaboration and competition \cite{biggerstaff2016results, biggerstaff2018results, mcgowan2019collaborative}. 
The FluSight challenge has been a leading driver of recent model development and flu forecasting advancements \cite{pei2018forecasting, kandula2018evaluation, pei2017counteracting, kandula2019reappraising, brooks2015flexible, farrow2017human, brooks2018nonmechanistic, osthus2017forecasting, osthus2019dynamic, osthus2019even, hickmann2015forecasting, ray2017infectious, ray2018prediction, reich2019collaborative, ben2018national, yang2015accurate, lu2019improved}.

Up until the 2016 flu season (i.e., the flu season starting in the fall of 2016 and ending in the spring of 2017), the FluSight challenge's scope encompassed forecasting short-term (one-to-four week ahead) and seasonal (season onset, peak timing, and peak intensity) targets at two geographic scales: nationally and regionally, where regions correspond to Health and Human Services (HHS) regions. 
National and regional forecasts give a high-level view of flu activity across the US. 
Those forecasts provide value to national and regional public health officials, but offer only coarse information for state and local public health practitioners. 
Thus, starting with the 2017 flu season, the FluSight challenge expanded to a third geographic scale: states and territories (referred to as states). 
This expansion to a finer geographic scale presents an opportunity to move forecasting to geographic scales better aligned with public health response infrastructure and decision making. 
It also presents an opportunity to develop and advance methodological forecasting frameworks that can share information across geographic locations, flu seasons, and geographic scales coherently in ways that geographically isolated forecasting models cannot.

\begin{figure}
\centering
\includegraphics[width=.8\linewidth]{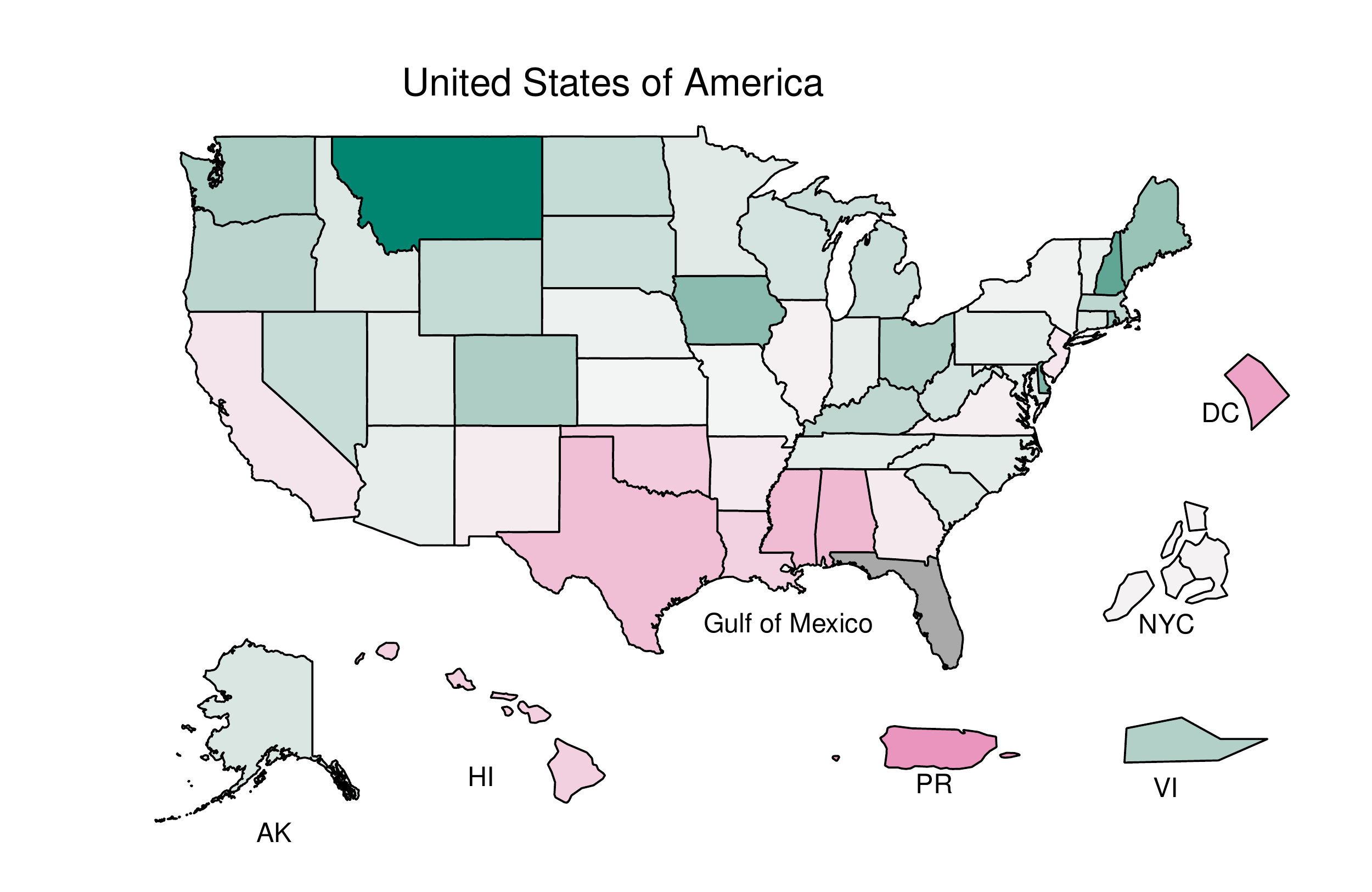}
\includegraphics[width=.8\linewidth]{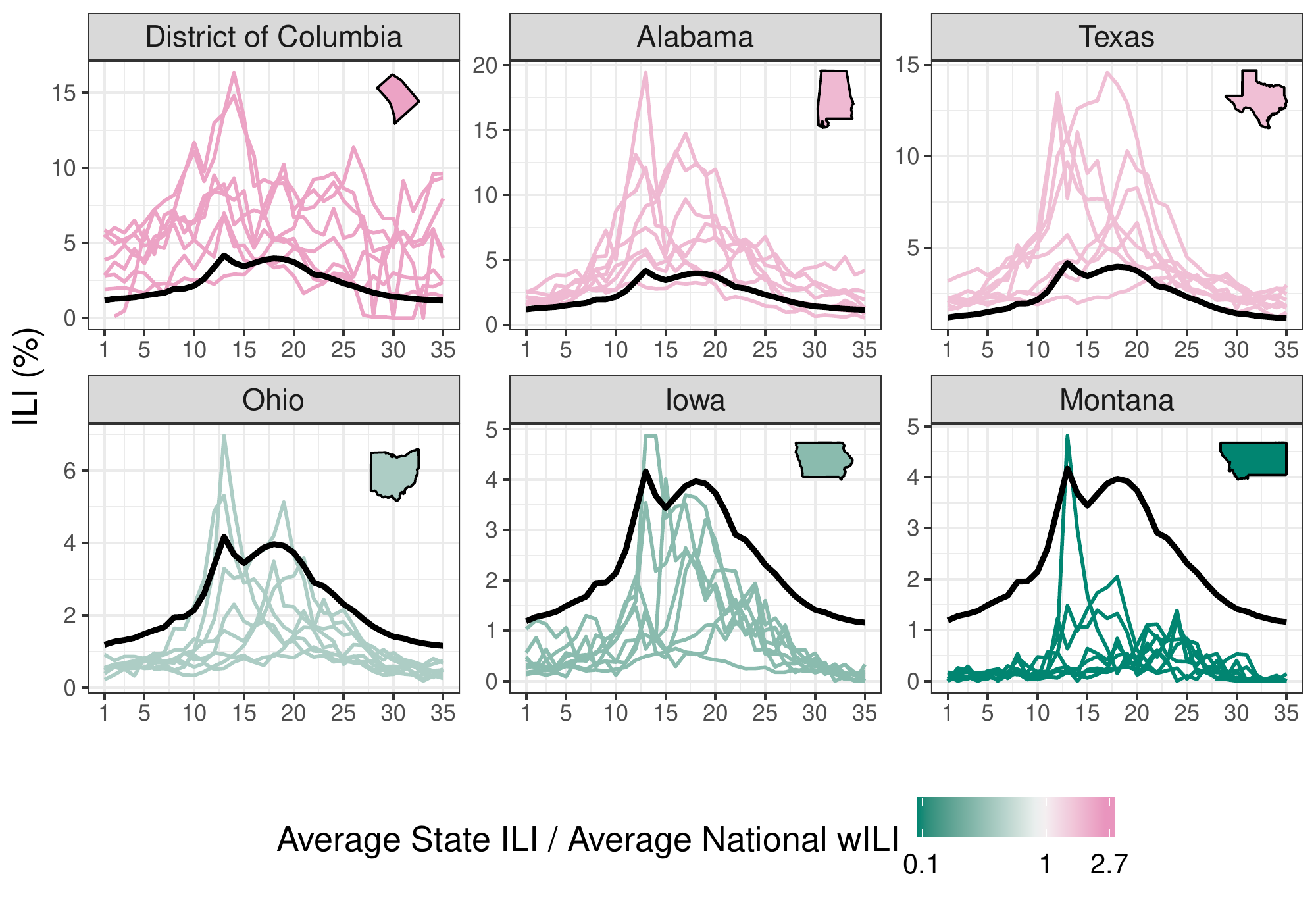}
\caption{(Top) Average state influenza-like illness (ILI) relative to average national weighted ILI (wILI). States bordering the Gulf of Mexico tend to have higher ILI than the national average. The geographical sizes of Alaska (AK), Hawaii (HI), Puerto Rico (PR), the US Virgin Islands (VI), New York City (NYC), and the District of Columbia (DC) are not to scale. Data for Florida is unavailable. Averages are based on 2010 through 2017 data. (Bottom) ILI by season (colored lines) for select states. Black line is national average wILI for reference. Appreciable season-to-season and state-to-state ILI variability exists.}
\label{fig:eda_state}
\end{figure}

Multiscale forecasting in the US requires careful consideration as it presents numerous challenges.
For instance, Figure \ref{fig:eda_state} (as well as SI Figure \ref{fig:eda_all_states}) shows appreciable state-to-state ILI variability. 
As an example, Montana's average ILI is about 20\% the national average, while the District of Columbia's and Puerto Rico's average ILI is about 250\% the national average. 
Figure \ref{fig:eda_state} also shows evidence of spatial correlation, with states near the Gulf of Mexico having higher than average ILI while most Midwest and Mountain West states have lower than average ILI.
There are, however, exceptions to these spatial trends with some higher than average ILI states exclusively surrounded by below average ILI states (e.g., California and New Jersey).
Attempts to model the spatial relationships of flu and flu-like illnesses include using network models and US commuter data \cite{pei2018forecasting}, network models based on Euclidean distance \cite{ben2018national}, and empirically derived network relationships \cite{lu2019improved,davidson2015using}. 
Though these approaches consider spatial relationships differently, they all support the conclusion that sharing information across geographies can improve forecasting.

\begin{figure}
\centering
\includegraphics[width=.45\linewidth]{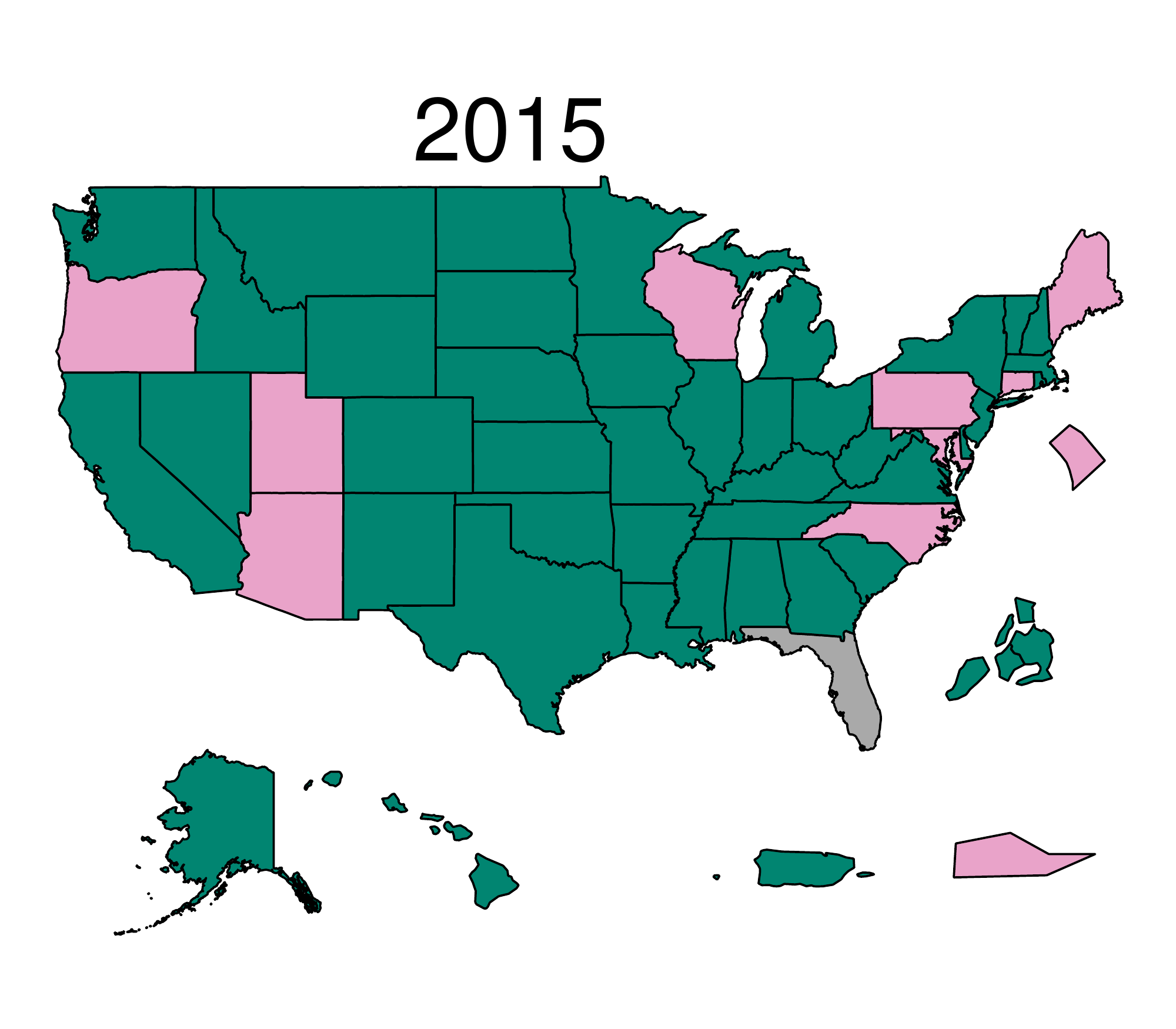}
\includegraphics[width=.45\linewidth]{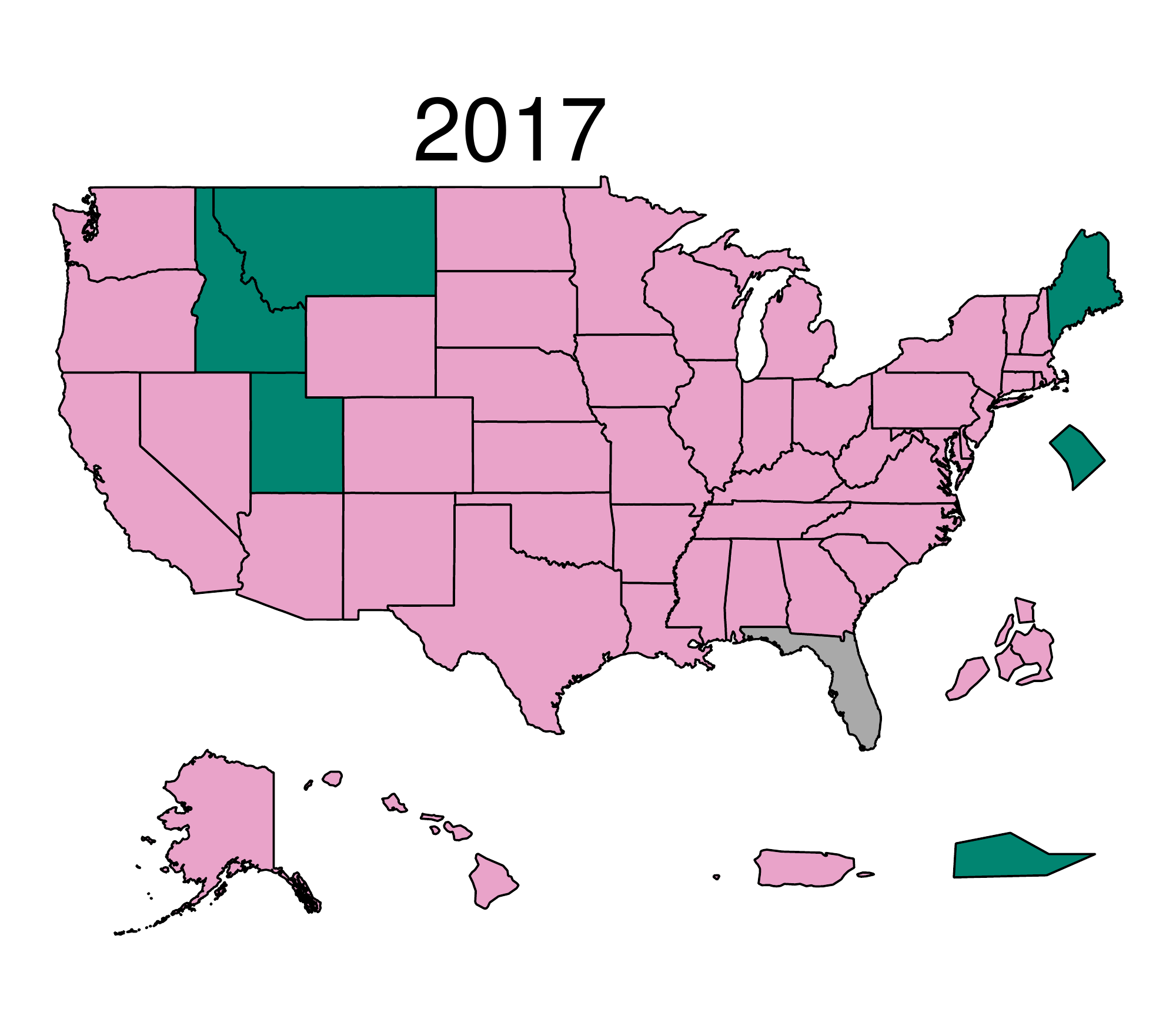}
\includegraphics[width=.45\linewidth]{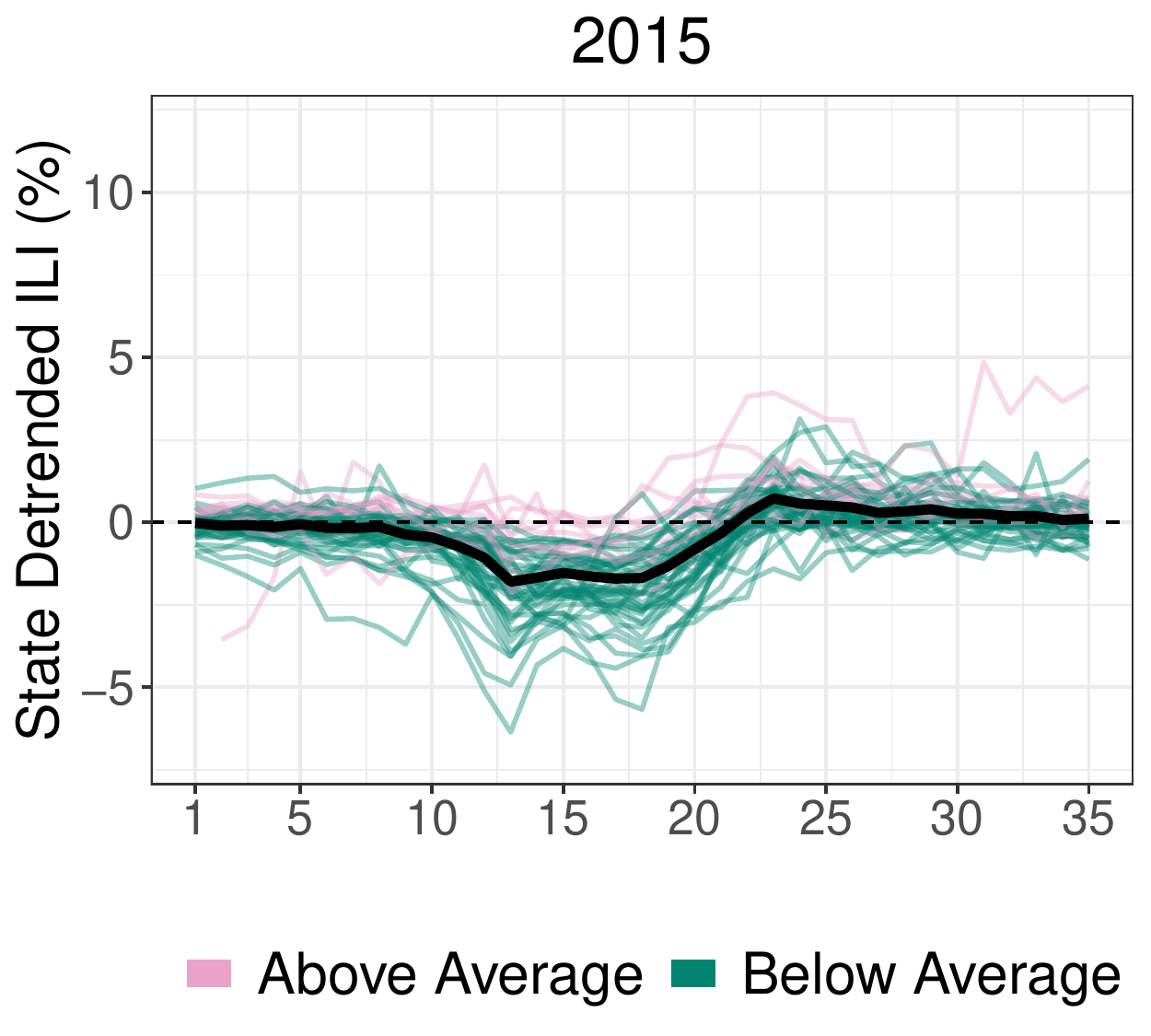}
\includegraphics[width=.45\linewidth]{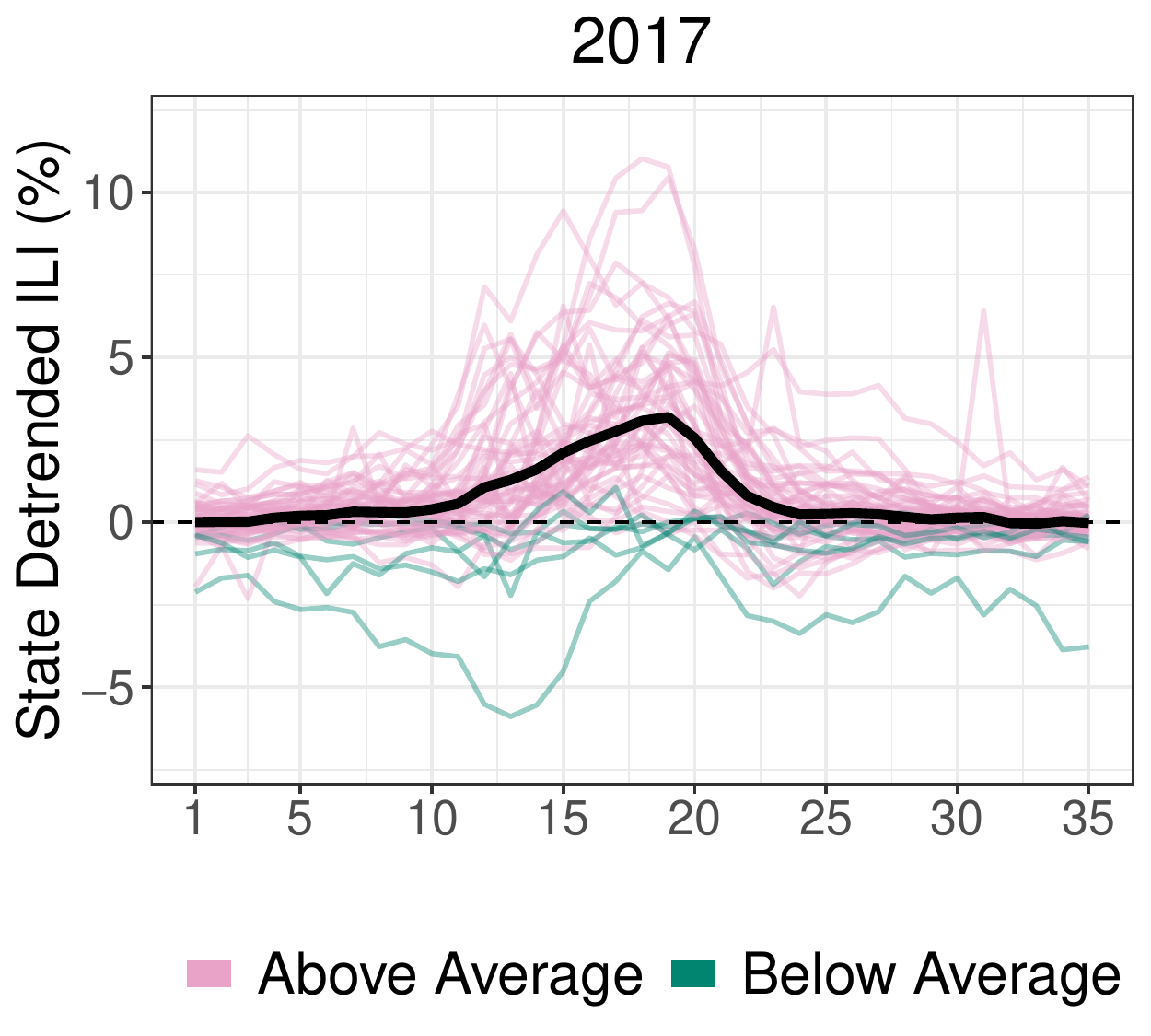}
\caption{(Top) Dark green states denote states with ILI less than their state-specific averages while pink states are states with ILI above their state-specific averages. 2015 was a mild flu season for the majority of states relative to their state-specific average ILI, while 2017 was an intense flu season for the majority of states, indicating that season-to-season effects can affect most of the country. Data unavailable for Florida. States displayed outside of the contiguous US are geographically not to scale. (Bottom) State detrended ILI for the 2015 and 2017 flu seasons, where state detrended ILI is ILI for a state/season minus ILI for that state averaged over all seasons. Positive/negative state detrended ILI means ILI for that season was above/below the state-specific average, respectively.}
\label{fig:eda_season}
\end{figure}

\begin{sidewaysfigure*}
	\centering
	\includegraphics[width=1\linewidth]{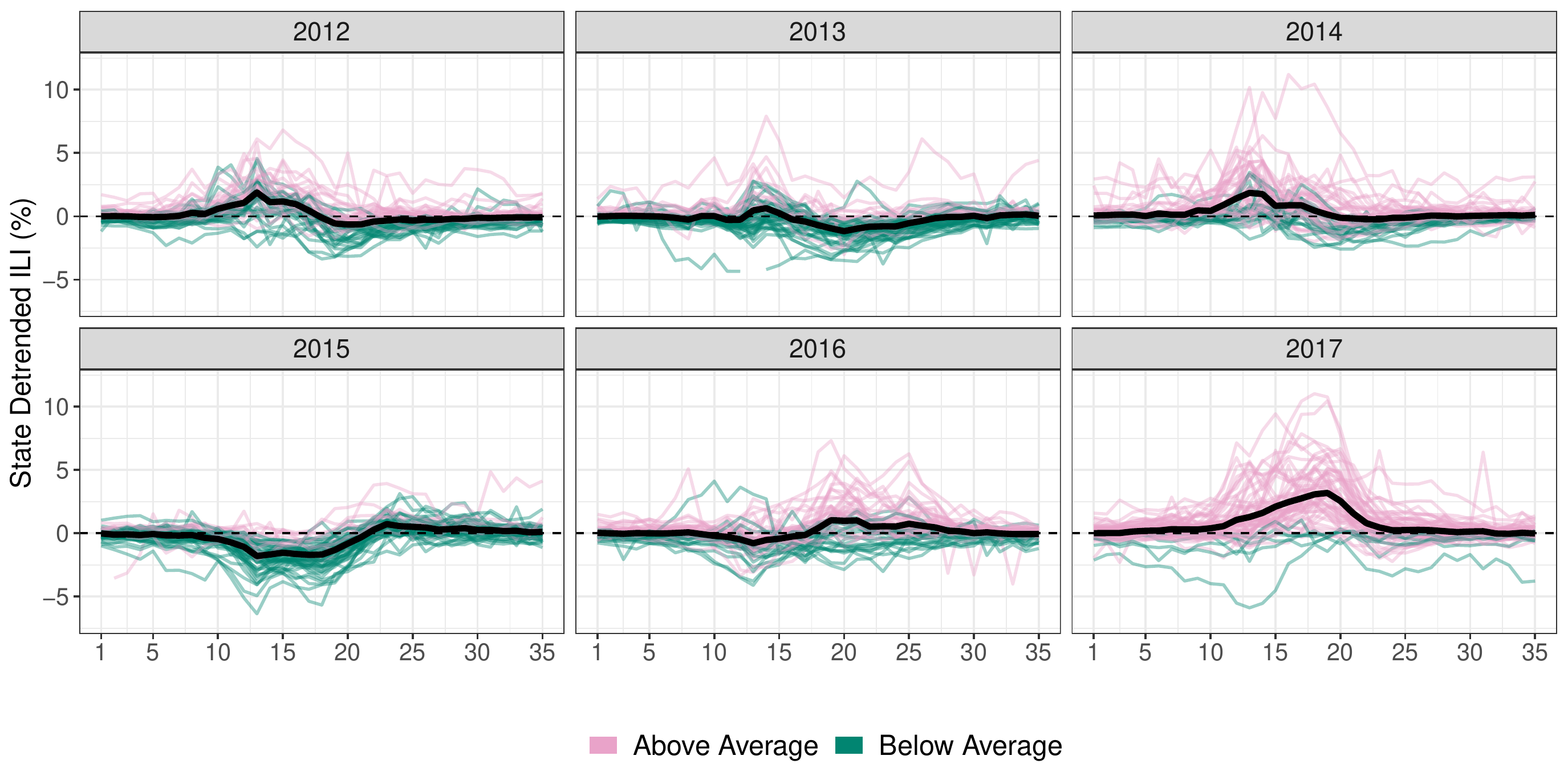}
	\caption{State detrended ILI time series where the state detrended time series is ILI for a given season minus the average ILI for the state averaged over all seasons. Pink/dark green lines correspond to seasons where ILI for that season was above/below its state-specific average, respectively. The black line for each season is the average state detrended ILI trajectory, averaged over all states within a season. 2015 was a mild season for the majority of states, while 2017 was an intense season for the majority of states.}
	\label{fig:eda_all_seasons}
\end{sidewaysfigure*}

Figure \ref{fig:eda_season} (and SI Figure \ref{fig:eda_all_seasons}) shows season-to-season variability, illustrating the common directional effect a flu season can have on nearly all states. 
2015, for instance, was a mild flu season in the US with 42 out of 53 states experiencing ILI activity below their state-specific averages (the 53 states are all 50 states, minus Florida with no available data, plus Puerto Rico, the US Virgin Islands, New York City, and the District of Columbia).
In contrast, 2017 was an intense flu season with 47 out of 53  states experiencing ILI activity above their state-specific averages. 
Similar to the findings that sharing information across geographies can improve flu forecasting, previous work has found that sharing information across seasons can also improve flu forecasting \cite{osthus2019dynamic}. 

\begin{figure}
\centering
\includegraphics[width=.45\linewidth]{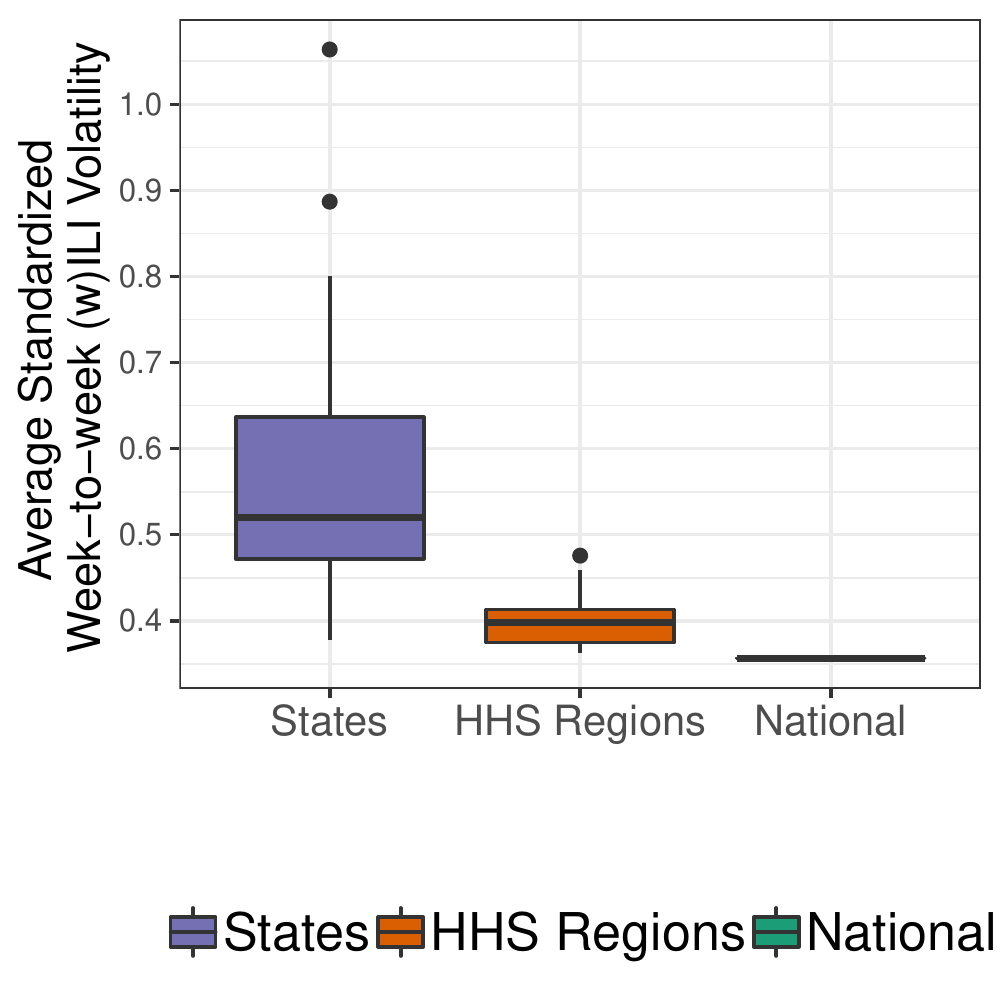}
\includegraphics[width=.45\linewidth]{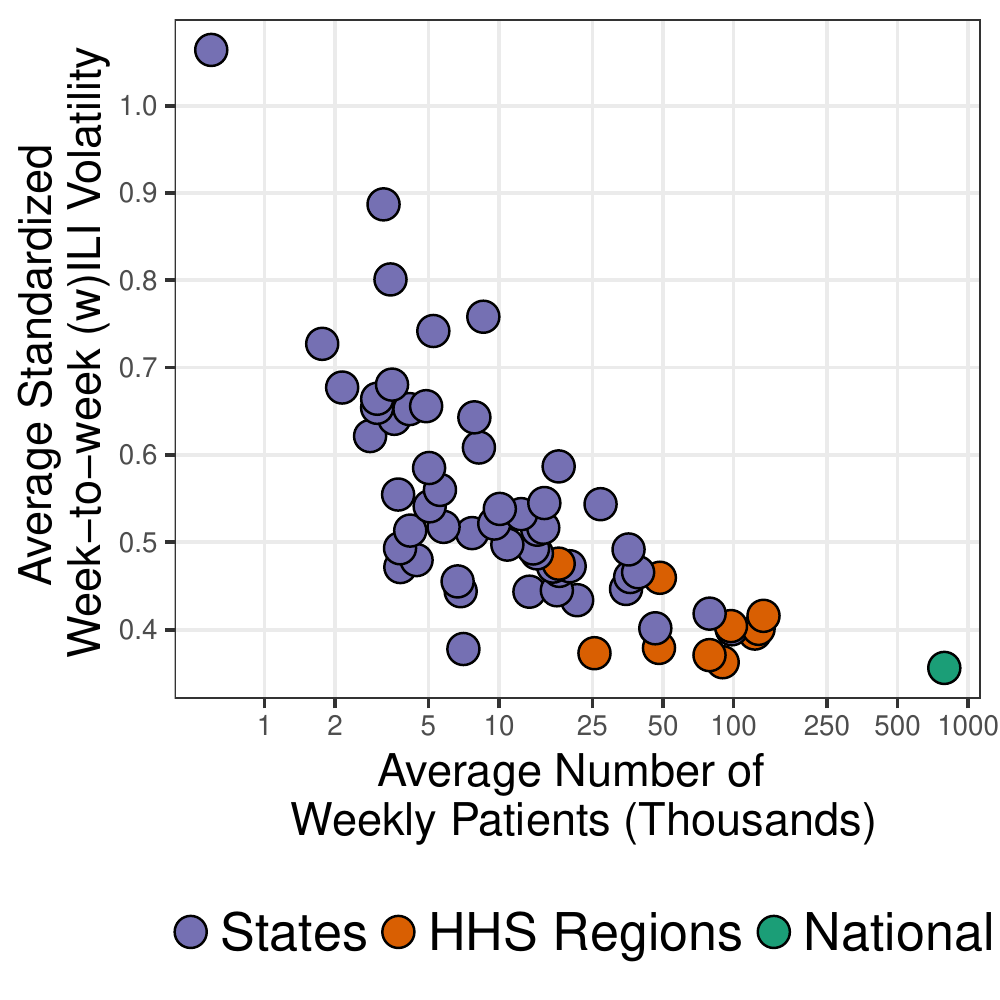}
\caption{(Left) Average standardized week-to-week influenza-like illness (ILI -- states) and weighted ILI (wILI -- HHS regions and nationally) volatility for three geographic scales. Volatility decreases as the scales coarsen. (Right) Average standardized week-to-week (w)ILI volatility versus the average number of weekly patients on a log scale for each state, HHS region, and nationally. Volatility decreases as the number of weekly patients seen increases, suggesting that volatility is in part a product of ILINet participation.}
\label{fig:eda_scale}
\end{figure}

Figure \ref{fig:eda_scale} shows the average standardized week-to-week volatility across geographic scales (see SI Section \ref{sec:standardized_volatility} for details). 
Standardized volatility measures how much ILI (states) and wILI (regions and nationally) varies from week-to-week, where wILI is weighted ILI \textemdash~a state-population weighted version of ILI used to characterize ILI regionally and nationally.
High volatility poses a challenge to forecasting as increased volatility can swamp the signal in the (w)ILI data.
Figure \ref{fig:eda_scale} makes clear that extending forecasts down to the state scale, a more actionable scale for public health officials, comes at the cost of increased volatility.
ILI is an estimate of the proportion of patients seen by ILINet with symptoms consistent with the flu relative to the total number of patients seen by ILINet for any reason. 
The level of noise in the estimates of those proportions are largely driven by the number of patients seen weekly. 
As illustrated in Figure \ref{fig:eda_scale}, the fewer patients seen weekly, the noisier the ILI estimates and the larger the volatility can be.
Thus, volatility will naturally increase with finer geographic scales, unless counterbalanced with increased ILINet participation.
Developing multiscale flu forecasting models that account for decreasing volatility with coarsening geographic scales will be crucial.
Some multiscale forecasting models have been developed in the context of norovirus gastroenteritis prediction \cite{held2017probabilistic}. 
In that work, \cite{held2017probabilistic} showed that modeling at the finest available data scale and aggregating up to coarser scales generally had better predictive performance than models directly operating at the aggregated scales.
To our knowledge, such models have not been developed and operationalized for ILI forecasting.

It is in the context of appreciable state-to-state and season-to-season variability, uneven ILINet surveillance participation, and the need to render short-term and seasonal probabilistic forecasts at nested geographic scales that Dante, a probabilistic, multiscale flu forecasting model, was developed.
While efforts have been made to address each of these challenges in isolation, no one has yet to tackle all of these challenges simultaneously in the context of influenza forecasting.
Jointly addressing all these challenges is the main contribution of this paper. 


\section*{The Dante Forecasting Model}
Dante is a probabilistic, Bayesian flu forecasting model that is decomposed into two submodels: the fine-scale model (i.e., the state model) and the aggregation model (i.e., the regional and national model).

\subsection*{Dante's Fine-Scale Model}
Dante's fine-scale model is itself described in two parts: the data model and the process model. 

\subsubsection*{Dante's Data Model}

Let $y_{rst} \in (0,1)$ be ILI/100 for state $r = 1,2,\ldots,R$ in flu season $s=1,2,\ldots,S$, for epidemic week $t=1,2,\ldots, T=35$, where $t=1$ corresponds to epidemic week 40, roughly the first week of October and $T=35$ most often corresponds to late May. 
Dante models the observed proportion $y_{rst}$ with a Beta distribution as follows:

\begin{align}
y_{rst}| \theta_{rst}, \lambda_r &\sim \text{Beta}(\lambda_r \theta_{rst}, \lambda_r (1 - \theta_{rst} ) ),
\end{align}
\noindent where

\begin{align}
\text{E}(y_{rst}| \theta_{rst}, \lambda_r) &= \theta_{rst}, \\
\text{Var}(y_{rst}| \theta_{rst}, \lambda_r) &= \frac{\theta_{rst} (1-\theta_{rst})}{1 + \lambda_r }. 
\end{align}

\noindent In Dante, $\theta_{rst}$ is the unobserved true proportion of visits for ILI in state $r$ for season $s$ during week $t$ and $\lambda_r > 0$ is a state-specific parameter that captures the level of noise in the ILINet surveillance system and thus the level of volatility in the ILI time series.
In Dante, $y_{rst}$ is modeled as unbiased for the latent state $\theta_{rst}$.
The observation $y_{rst}$, however, is not equal to $\theta_{rst}$ due to variability in the measurement surveillance process (i.e., the true proportion of ILI in state $r$ for season $s$ during week $t$ is not going to be perfectly captured by ILINet surveillance).
Motivated by Figure \ref{fig:eda_scale}$, \lambda_r$ is likely to be related to ILINet participation as measured by the total number of patients seen weekly in state $r$.
As $\lambda_r$ increases, the variance of $y_{rst}$ decreases and observations will tend to be closer to $\theta_{rst}$.
Because we do not know the relationship between patient count and $\lambda_r$ \emph{a priori}, we model $\lambda_r$ hierarchically, allowing them to be learned from data (details in SI Section \ref{sec:data_model}).

\begin{figure}
\centering
\includegraphics[width=.6\linewidth]{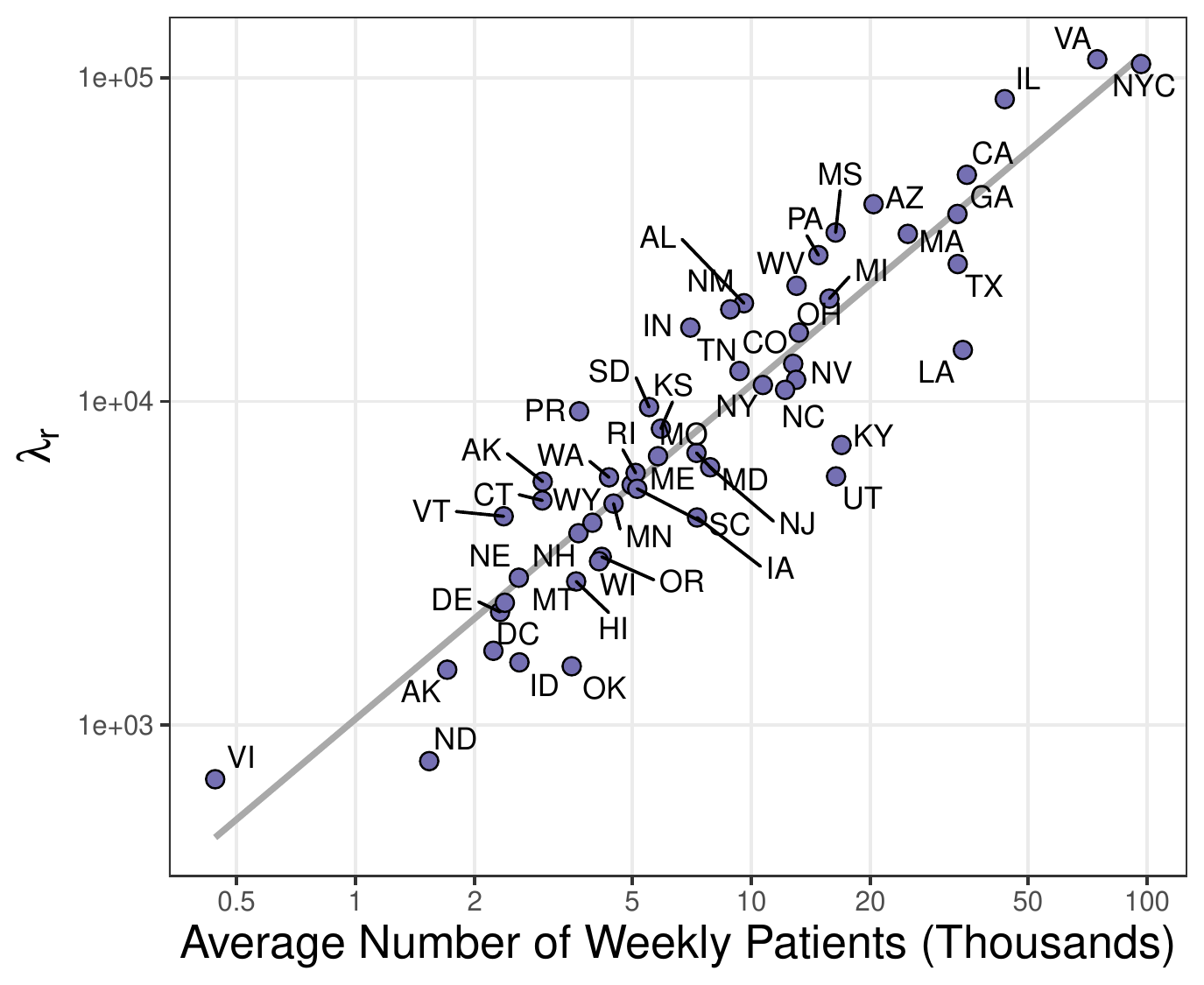}
\caption{The posterior mean for $\lambda_r$ versus the average number of patients seen weekly by each state. Both axes are on a log scale. A clear linear relationship is observed. Dante learns this relationship, as it has no explicit knowledge of the average number of patients.}
\label{fig:results_lambda_postpred}
\end{figure}

Figure \ref{fig:results_lambda_postpred} shows the posterior mean of $\lambda_r$ versus the average number of patients seen weekly by each state. 
A clear linear relationship is observed on a log-log scale, where the variance of $y_{rst}$ goes down (i.e., $\lambda_r$ increases) as the total weekly seen patients increases.
What is particularly striking about Figure \ref{fig:results_lambda_postpred} is that Dante has no knowledge of the number of patients seen each week as it is not an input to Dante, illustrating how structure can be learned rather than prescribed with a flexible, hierarchical model.

\subsubsection*{Dante's Process Model}
Dante's process model models the unobserved true proportion of ILI, $\theta_{rst} \in (0,1)$, as a function of four components:

\begin{align}
\theta_{rst} &= \text{logit}^{-1}(\pi_{rst}), \label{eq:theta}\\
\pi_{rst} &= \mu^{\text{all}}_t + \mu^{\text{state}}_{rt} + \mu^{\text{season}}_{st} + \mu^{\text{interaction}}_{rst}. \label{eq:pi}
\end{align}
\noindent The four terms in Equation \ref{eq:pi} are modeled as random or reverse-random walks, allowing patterns in the process model to be flexibly learned while capturing week-to-week correlation (details in SI Section \ref{sec:process_model}).

\begin{figure*}[tbhp]
\centering
\includegraphics[width=1\linewidth]{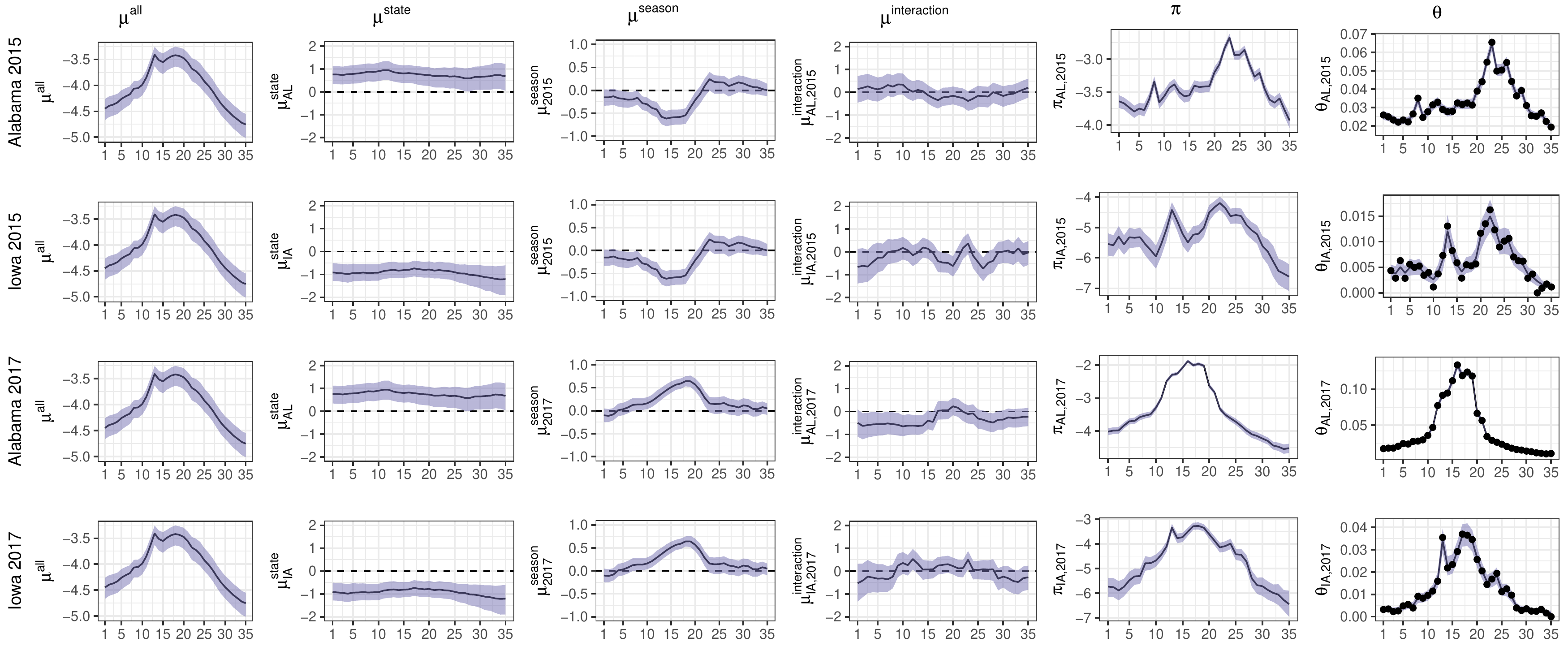}
\caption{Posterior summaries for select components, seasons, and states of Dante, fit to seasons 2010 through 2017. Rows, from top to bottom, correspond to Alabama in 2015, Iowa in 2015, Alabama in 2017, and Iowa in 2017. Columns, from left to right, correspond to $\mu^{\text{all}}$, $\mu^{\text{state}}$, $\mu^{\text{season}}$, $\mu^{\text{interaction}}$, $\pi$ (all from Equation \ref{eq:pi}), and $\theta$ (Equation \ref{eq:theta}). The $\pi$ column is the sum of the $\mu^{\text{all}}$, $\mu^{\text{state}}$, $\mu^{\text{season}}$, and $\mu^{\text{interaction}}$ columns, accounting for posterior covariances. The $\theta$ column is the inverse logit of the $\pi$ column and is back on the scale of the data. The $\mu^{\text{all}}$ component is the most structured component, as it is common for all states and seasons (i.e., the same for all rows). The components $\mu^{\text{state}}$ and $\mu^{\text{season}}$ are the next most structured components. They describe the state-specific and season-specific deviations from $\mu^{\text{all}}$, respectively, and are common for all seasons within a state ($\mu^{\text{state}})$ and all states within a season ($\mu^{\text{season}}$). The component $\mu^{\text{interaction}}$ is the least structured component of Dante, as it is specific to each season/state (i.e., it is different for each row).  Solid lines are posterior means. Ribbons are 95\% posterior intervals. In the $\theta$ column, points are data, $y$. }
\label{fig:results_example_postpred}
\end{figure*}

Figure \ref{fig:results_example_postpred} illustrates the fits for all model components for Alabama and Iowa for the 2015 and 2017 flu seasons.
The component $\mu^{\text{all}}_t$ is common to every state and season and acts as the anchor for the process model.
The shape of $\mu_t^{\text{all}}$ is similar to the national average ILI trajectory of Figure \ref{fig:eda_state}, capturing the profile for a typical state and season.
The component $\mu^{\text{state}}_{rt}$ captures the state-specific deviation from $\mu^{\text{all}}_t$ and is common to every season for a given state, but is distinct for each state.
As can be see in Figure \ref{fig:eda_state}, Alabama typically sees ILI above the national average, hence why $\mu^{\text{state}}_{rt}$ for Alabama is learned to be greater than zero. 
Iowa, however, typically sees ILI below the national average, explaining why $\mu^{\text{state}}_{rt}$ is learned to be less than zero for Iowa.
The component $\mu^{\text{season}}_{st}$ captures the season-specific deviation from $\mu^{\text{all}}_t$ and is common to every state for a given season, but is distinct for each season.
This component captures the fact that seasons can have effects that are shared by nearly all states, as illustrated in Figure \ref{fig:eda_season}.
The shape of $\mu^{\text{season}}_{st}$ for 2015 and 2017 has a similar shape to the average residuals for 2015 and 2017, respectively, in Figure \ref{fig:eda_season}.
Finally, $\mu^{\text{interaction}}_{rst}$ captures the remaining signal in $\pi_{rst}$ that cannot be accounted for by $\mu^{\text{all}}_t$, $\mu^{\text{state}}_{rt}$, and $\mu^{\text{season}}_{st}$.
The term $\mu^{\text{interaction}}_{rst}$ is distinct for each state and season.

Dante's process model is purposely over-specified. If our interest were purely to \textit{fit} ILI data, the term $\mu^{\text{interaction}}_{rst}$ alone would suffice. 
However, there is not enough structure to \textit{forecast} effectively with only $\mu^{\text{interaction}}_{rst}$.
On the other hand, the non-interaction terms in the decomposition of $\pi_{rst}$ ($\mu^{\text{all}}_{t}$, $\mu^{\text{state}}_{rt}$, and $\mu^{\text{season}}_{st}$) provide structure for forecasting but not enough flexibility to capture all the signal in the ILI data.
Thus, the $\mu^{\text{interaction}}_{rst}$ term provides the flexibility needed to fit the data, but is specified so that it plays as minimal a role as possible so that signal is captured in the non-interaction terms and can drive the shape of forecasts.

Inference for unobserved components of Dante, as well as state-level forecasts of yet-to-be-observed $y_{rst}$ are generated by sampling from the posterior distribution with Markov chain Monte Carlo (MCMC), resulting in a sample of $M$ draws that summarize the posterior distribution (details in SI Section \ref{sec:mcmc}).
We use the software JAGS (Just Another Gibbs Sampler ) \cite{plummer2003jags}, as called by the \texttt{R} package \texttt{rjags} \cite{rjags2018} within the programming language \texttt{R} \cite{r2018} to perform the MCMC sampling.
We denote each MCMC draw by the index $m$. Notationally, we denote the $m^{th}$ sample for a yet-to-be-observed $y_{rst}$ as $y_{rstm}$.

\subsection*{Dante's Aggregation Model}
Dante's regional and national forecasts are computed as linear combinations of state forecasts, where weights are proportional to 2010 US Census population estimates (SI Figure \ref{fig:aggregation}).
Let $w_r \in [0,1]$ be the population of state $r$ divided by the US population such that $\sum^R_{r=1} w_r = 1$. 
For each MCMC draw $m$, we compute the ILI forecast for aggregated region $\rho$ (indexing all ten HHS regions and nationally) as:

\begin{align}
y_{\rho stm} &= \sum_{r = 1}^R w^{\rho}_r y_{rstm}.
\end{align}
For $\rho =$ region X (e.g., $\rho = $ HHS Region 1), $w^{\rho}_r = 0$ if state $r$ is not a member of region X and 

\begin{align}
w^{\rho}_r &= \frac{w_r \text{I}(r \in \text{region X})}{\sum_{r=1}^R w_r \text{I}(r \in \text{region X})}
\end{align}
\noindent and $\text{I}(r \in \text{region X})$ is an indicator function equal to 1 if state $r$ is in region X and 0 otherwise.
By construction, $\sum_{r=1}^R w^{\rho}_r = 1$ for any $\rho$.
The aggregation model constitutes a bottom-up forecasting procedure and ensures forecasts are consistent across scales.

\section*{Results}
Dante is compared to a leading flu forecasting model, the Dynamic Bayesian Model (DBM).
DBM combines a susceptible-infectious-recovered (SIR) compartmental model with a flexible statistical discrepancy function to forecast ILI (see \cite{osthus2019dynamic} for more details).
DBM is fit to each geographic unit separately, thus does not share information across geographic units or geographic scales, but does share information across flu seasons.
In contrast, Dante shares information across geographic units, geographic scales, and flu seasons.
DBM was the 4$^{\text{th}}$ place model and a component model in the 2$^{\text{nd}}$ place ensemble model \cite{reich2019performance} in the prospective national and regional 2017/18 FluSight challenge out of 29 participating models.
Dante was the 1$^{\text{st}}$ place model in the 2018/19 national and regional FluSight challenge out of 33 participating models.
Dante also came in 1$^{\text{st}}$ out of 14 competing models in the 2018/19 state challenge.

We compare Dante and DBM using forecast skill following the scoring rules of the CDC's FluSight challenge (details in SI Section \ref{sec_main:Scoring_proc}), noting that this scoring rule is improper \cite{reich2019collaborative, bracher2019mblogscore, reich2019mblogscore}.
Forecast skill ranges between 0 and 1, with 1 being the best possible forecast skill.
Conceptually, skill is a function of both accuracy (a measure of point-estimates) and confidence (a measure of distributional sharpness).
We will show how Dante compares to DBM broadly in terms of skill, and also in terms of its component pieces.
Both models were fit in a leave-one-season out fashion, where the data for all seasons not being forecasted along with the data for the season being forecasted up to the forecast date were used for training.

Table \ref{tab:results_scales} shows that Dante outperformed DBM in forecast skill at all geographic scales. 
For both models, forecast skill improves as geographic scales coarsen, suggesting that forecast skill degrades as we move to finer scale geographies where volatility is greater.

\begin{table}[ht]
\centering
\caption{Dante and DBM average forecast skill comparisons across geographic scales. Forecast skill improves for both models as the geographic scale coarsens. Dante outperformed DBM at all geographic scales.}
\label{tab:results_scales}
\begin{tabular}{lrrr}
  \hline
\textbf{Model} & \textbf{States} & \textbf{HHS Regions} & \textbf{National} \\ 
  \hline
Dante & 0.372 & 0.413 & 0.439 \\ 
DBM & 0.337 & 0.383 & 0.426 \\ 
   \hline
\end{tabular}
\end{table}

Table \ref{tab:results_scales_mse} shows that Dante outperformed DBM in terms of accuracy, as measured by mean squared error (MSE) of point predictions, at all geographic scales. 
For both models, average MSE decreases as geographic scales coarsen, suggesting that accuracy degrades as we move to finer scale geographies where volatility is greater.
See SI Section \ref{sec:MSE} for further details and figures comparing the MSE of Dante and DBM.

\begin{table}[ht]
	\centering
	\caption{Dante and DBM average mean squared error (MSE) comparisons across geographic scales. MSE improves for both models as the geographic scale coarsens. Dante outperformed DBM at all geographic scales.}
	\label{tab:results_scales_mse}
	\begin{tabular}{lrrr}
		\hline
		\textbf{Model} & \textbf{States} & \textbf{HHS Regions} & \textbf{National} \\ 
		\hline
		Dante & 3.164 & 2.441 & 1.921 \\ 
		DBM & 3.390 & 2.674 & 2.244 \\ 
		\hline
	\end{tabular}
\end{table}

\begin{figure}[tbhp]
\centering
\includegraphics[width=.7\linewidth]{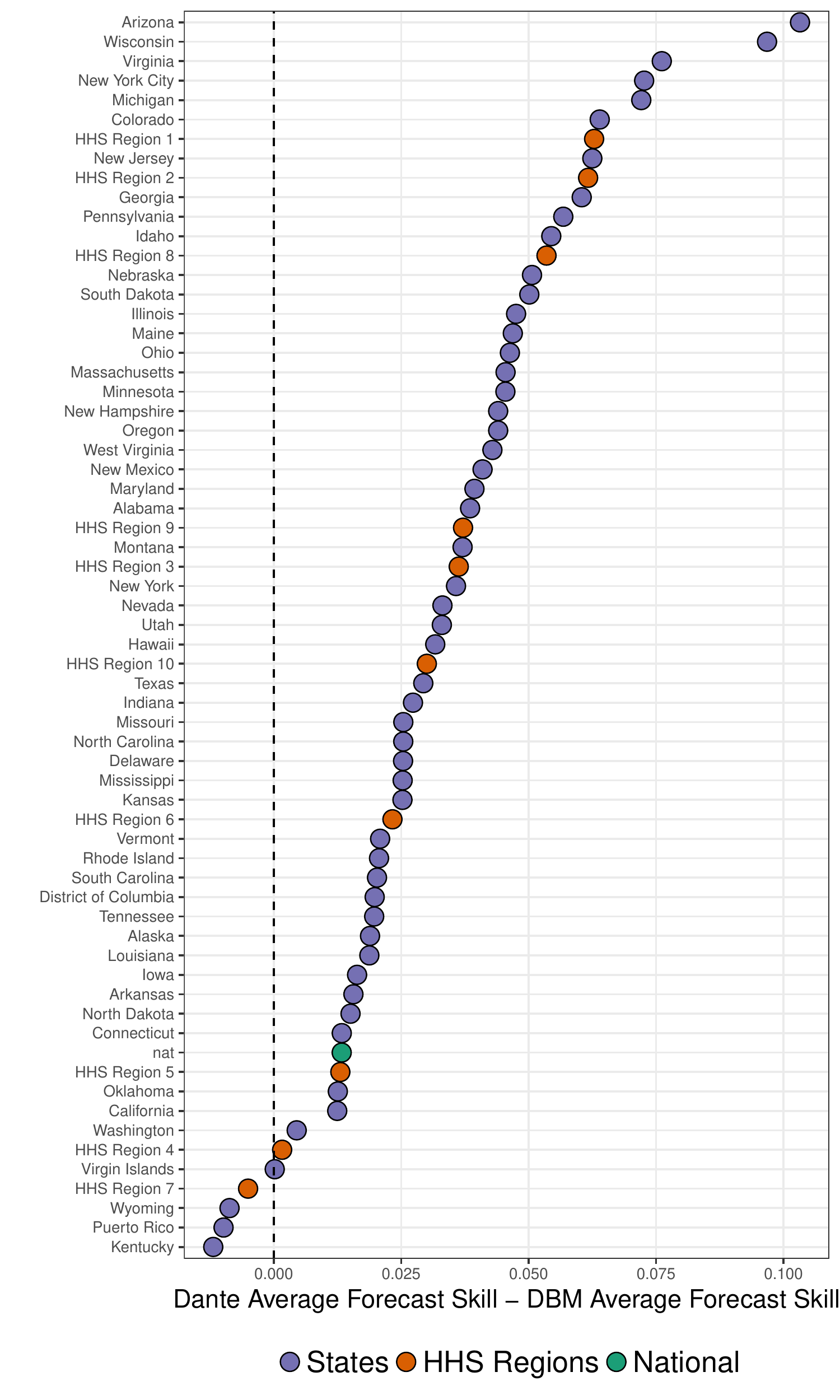}
\caption{Difference in forecast skill between Dante and DBM, for all states, regions, and nationally. Dante had higher forecast skill for all geographic regions except for HHS Region 7, Kentucky, Wyoming, and Puerto Rico.}
\label{fig:results_regions}
\end{figure}

Figure \ref{fig:results_regions} shows the difference in forecast skill between Dante and DBM for each state, region, and nationally. 
Dante outperformed DBM for the majority of geographic regions, with the exception of HHS Region 7 and the states Wyoming, Puerto Rico, and Kentucky.

\begin{figure}[tbhp]
\centering
\includegraphics[width=.6\linewidth]{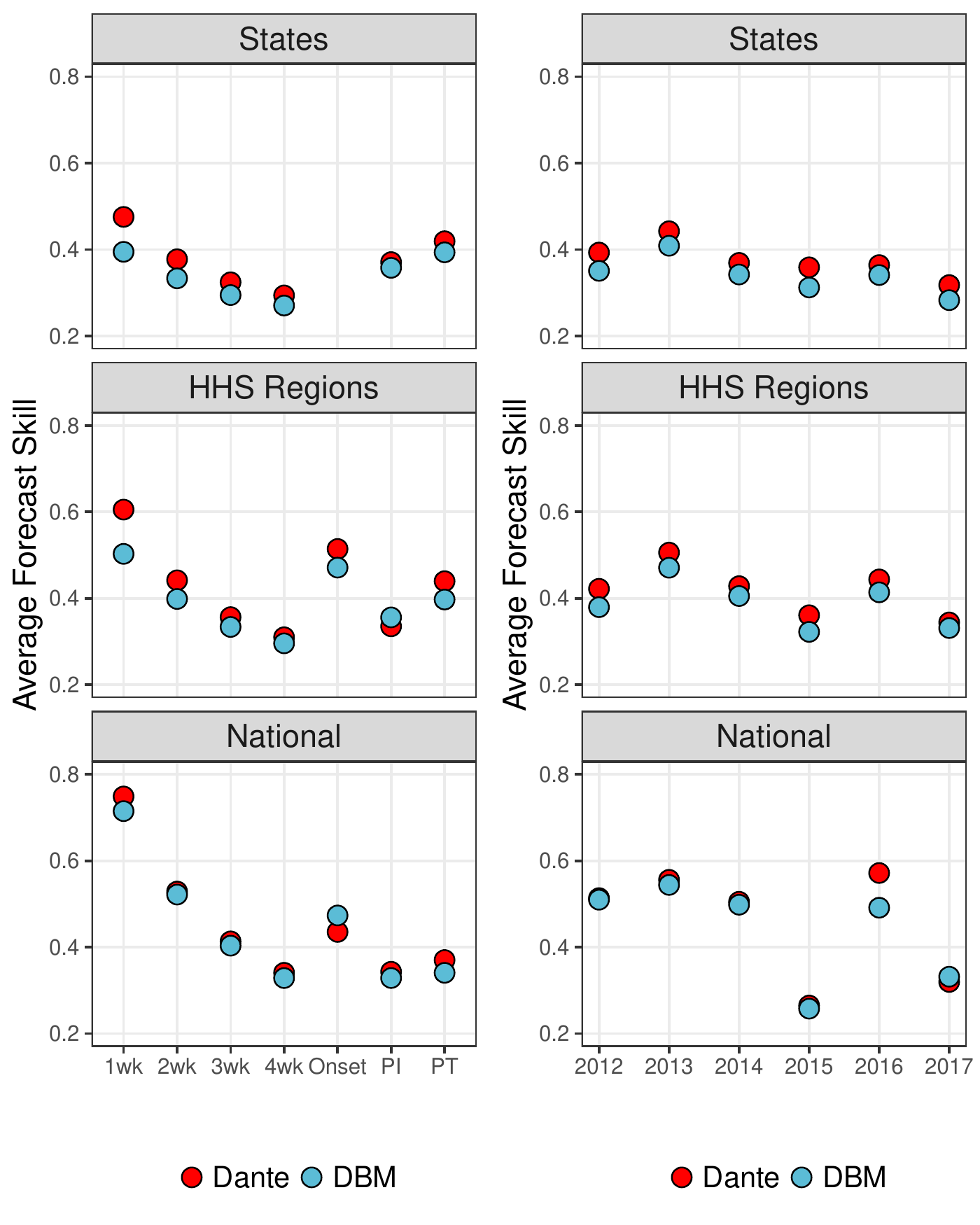}
\caption{(left) Average forecast skill by scales and targets. PI and PT stand for peak intensity and peak timing, respectively. Dante outperformed DBM for all scales and targets, except for onset nationally and for peak intensity (PI) regionally. (right) Average forecast skill by scales and flu seasons. Dante outperformed DBM for all scales and targets, except for 2017 nationally.}
\label{fig:results_targets_and_seasons}
\end{figure}

Figure \ref{fig:results_targets_and_seasons} shows forecast skill broken down by targets (left) and flu seasons (right) for each geographic scale.
Dante outperformed DBM for all scales and targets, except for peak intensity regionally and onset nationally.
Improvement over DBM is largest for the 1-week ahead forecast target.
Dante also outperformed DBM for all scales and flu seasons, except for 2017 nationally.
While forecast skill for DBM and Dante are close for all seasons nationally (sans 2016), Dante consistently and appreciably outperformed DBM for all seasons at the regional and state scales.

\begin{figure}[tbhp]
\centering
\includegraphics[width=.7\linewidth]{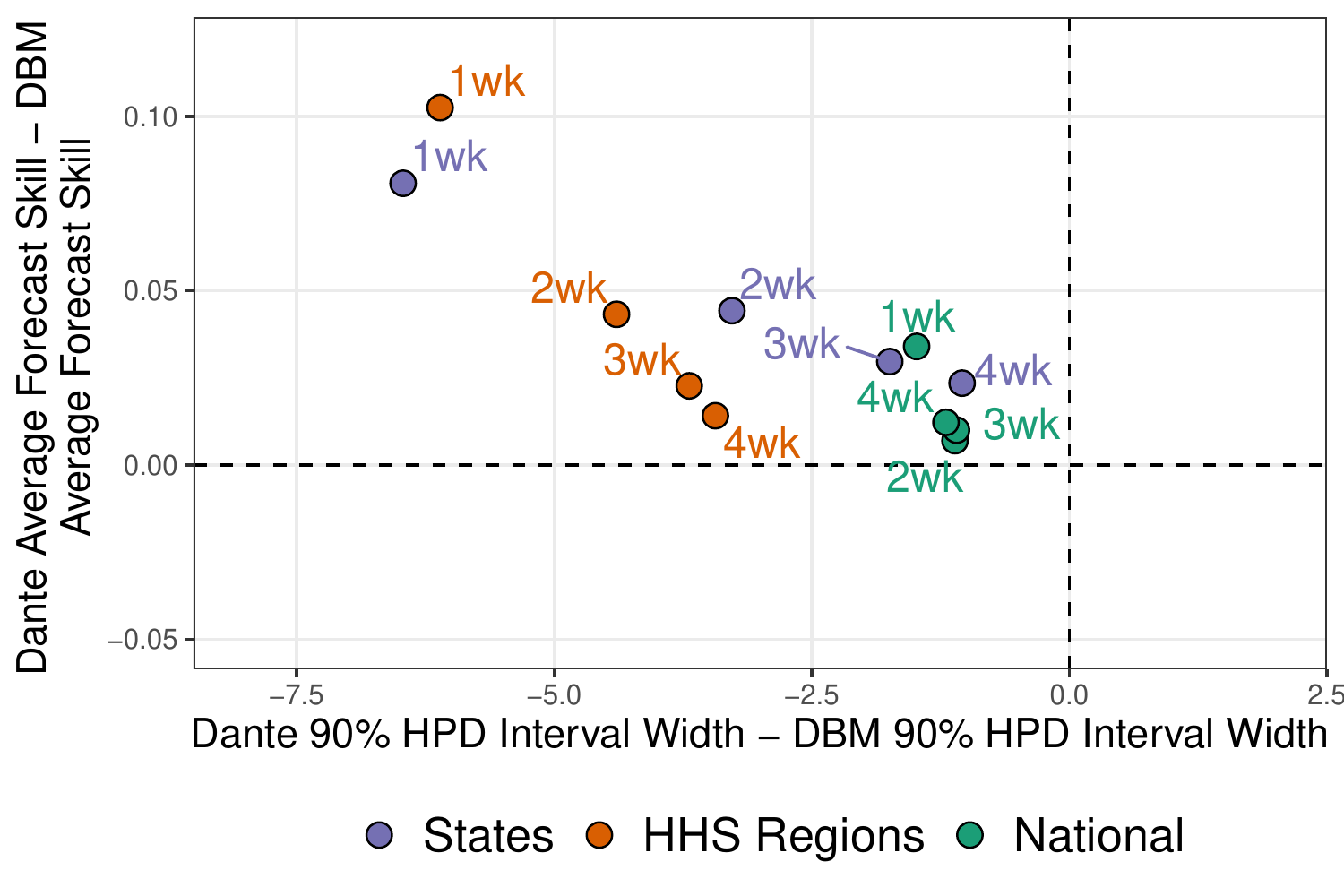}
\caption{Difference in average forecast skill versus difference in 90\% highest posterior density (HPD) interval widths for short-term forecasting targets. For all short-term forecasting targets and geographic scales, Dante produced more confident (i.e., smaller 90\% HPD interval widths) and higher scoring forecasts than DBM.}
\label{fig:hpd}
\end{figure}

Figure \ref{fig:hpd} provides context as to how Dante is outperforming DBM. 
Figure \ref{fig:hpd} displays the difference in forecast skill at each scale for all short-term forecasts against the difference in the 90\% highest posterior density (HPD) widths for each of the short-term target's posterior predictive distributions. 
HPDs are similar to confidence intervals as they capture the range of probability concentration, but are more appropriate than confidence intervals for distributions that are not unimodal.
Figure \ref{fig:hpd} shows that for all scales and short-term forecasts, Dante has smaller HPD interval widths, indicating that Dante's forecasts are more concentrated than DBM and thus, more confident.
Dante's increased forecast confidence resulted in higher forecast skill than DBM. 
This is a promising finding, as more confident forecasts, if accurate, provide more information to public health decision makers.

\begin{figure}[tbhp]
\centering
\includegraphics[width=.5\linewidth]{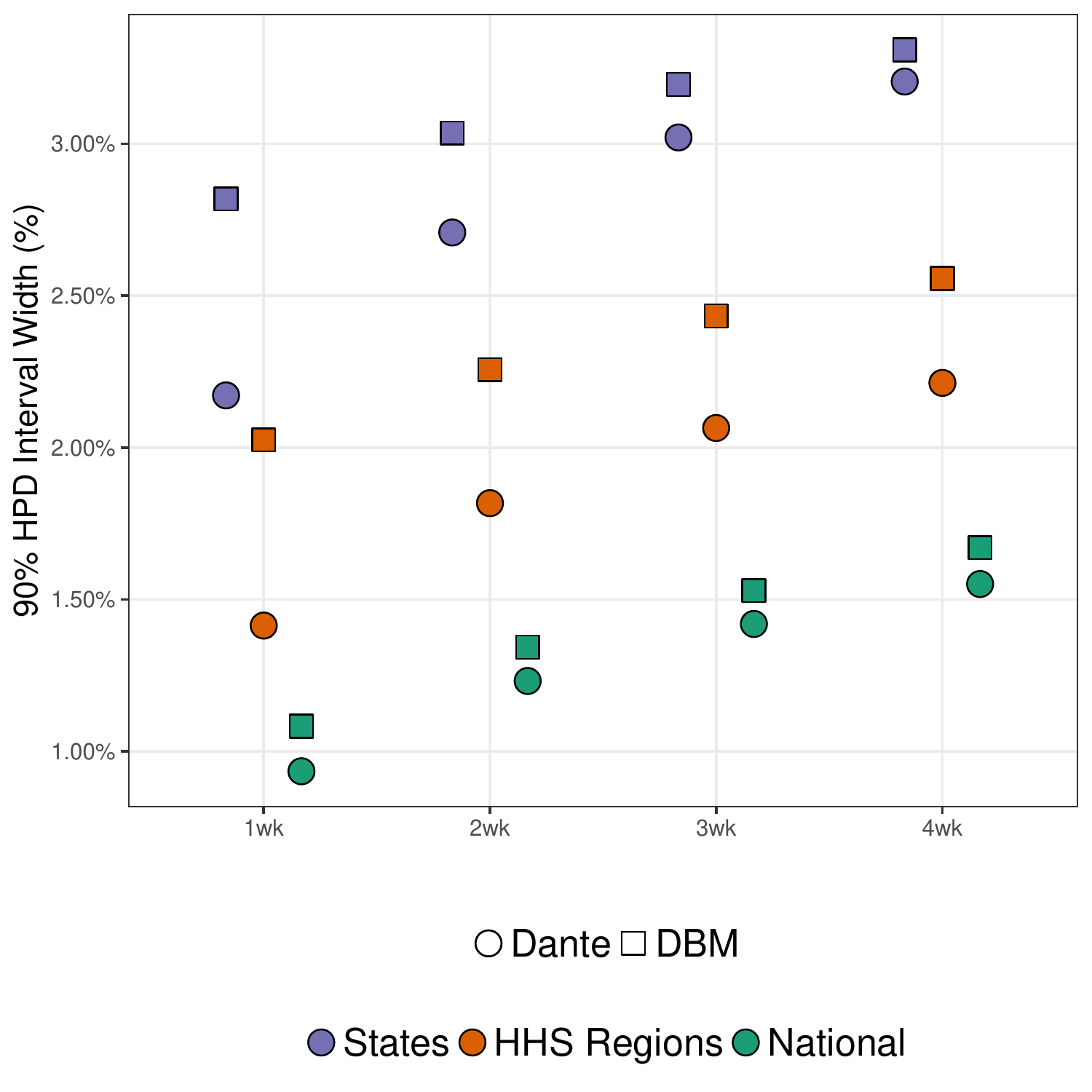}
\caption{Average 90\% highest posterior density (HPD) interval widths for the short-term forecasting targets. Both Dante and DBM produce more confident (i.e., smaller 90\% HPD interval widths) forecasts for coarser geographic resolutions. For all short-term forecasting targets and geographic scales, Dante produced more confident forecasts than DBM. }
\label{fig:hpd_scales}
\end{figure}

Figure \ref{fig:hpd_scales} shows that Dante's forecasts are more confident than DBM's for all short-term targets at all geographic scales. 
For each short-term target, forecasts for both DBM and Dante become more confident as the geographic scale coarsens.
DBM makes more confident short-term forecasts because the (w)ILI DBM is modeling is less volatile, i.e. because (w)ILI becomes less volatile as geographic scales coarsen.
Dante makes more confident short-term forecasts at coarsening geographic scales as a result of the aggregation model.
Dante's 3-week-ahead 90\% HPD interval widths nationally and regionally are 1.4\% and 2.1\%, respectively, about the same as Dante's 1-week-ahead 90\% HPD interval widths are regionally and at the state-level, respectively.
Said another way, Dante loses about 2 weeks of confidence in its short-term forecasts for each disaggregating geographic scale.

\subsection*{Discussion}

The plot of skill score (Figure \ref{fig:results_targets_and_seasons}) and total patients seen and volatility (Figure \ref{fig:eda_scale}) suggests that Dante (and DBM) can forecast geographic regions better when the forecasted estimate is based on more data than less. This result suggests that expanded ILI surveillance participation plays a role in improving model forecasts, not just improvements to the models themselves. This idea is not surprising. Disease forecasting has been compared to weather forecasting \cite{moran2016epidemic}, a field that has continued to make consistent progress through parallel efforts of improved modeling and data collection.

We found that Dante made more confident (or sharper) forecasts, as measured by smaller 90\% HPD interval widths than DBM, a model that fits aggregate data directly. 
Similar findings were noted by \cite{held2017probabilistic} when comparing models fit to norovirus data in Berlin stratified by regions and age groups \textemdash~models fit to the finest scales and subsequently upscaled had sharper predictions than models fit to the aggregated scales directly. 
This suggests that continued stratification of ILI, such as partitioning state-level ILI by age groups, by flu strain type, or county-level, may provide further sharpening of forecasts at aggregate scales.

Dante's first place finish in the 2018/19 FluSight challenge may come as a surprise given that it is a purely statistical model and uses only ILINet data, while many of its competitors are based in full or in part on mechanistic disease transmission models and/or are augmented with alternative data sources (e.g., Google search data). Dante's superior performance suggests that these mechanistic components or alternative data sources may be integrated into those models in a way that is improperly aligned with the truth. For example, DBM includes an SIR model component via a season-specific ``I'' term but a given season may have multiple circulating influenza strains responsible for the ``true'' flu component in the ILI data, thus rendering the use of a single ``I'' term inappropriate. 

Revisions made to ILI data after its initial release are referred to as backfill and constitute a meaningful source of uncertainty when prospectively forecasting ILI. In this work, backfill was ignored for the forecasting of both Dante and DBM. As a result, the forecasting results for Dante and DBM are directly comparable to each other in this paper but are not directly comparable to previous FluSight challenge results. The reason backfill was ignored in this paper is because Dante uses state-level ILI data directly, and state-level backfill data is only available starting for the 2017/18 season. Though backfill was not addressed in this paper, Dante's winning 2018/19 FluSight challenge entry did include a backfill model to account for the revision process of real-time ILI data.

Linking understandable processes to observed patterns in the data via models while maintaining high performance is the next frontier in ILI modeling. To do so will require a fuller consideration of the ``ILI data generating process.'' This process non-exhaustively includes a disease transmission process(es) (e.g., things often modeled with a compartmental model(s)), a healthcare visitation process (i.e., a set of processes related to who interacts with the healthcare system and when), an ILINet participation process (i.e., certain providers participate in ILINet while others do not, and the composition of provider networks varies temporally and spatially), and a reporting process (e.g., backfill). 

Further stratification is a promising direction for incorporating known facets of the ILI data generating process into Dante in a flexible way. For example, if provider-level ILINet data were available state-level models could be decomposed into models for emergency department (ED) ILI and non-ED ILI. We hypothesize that a systematic difference exists between patients visiting ED and non-ED providers, specifically that the proportion of ED patients with ILI is higher than that of non-ED patients. If so, then provider composition could help explain some state-level variation in ILI magnitude (we expect states with more ED providers have higher reported ILI). It could also explain part of the holiday-specific spikes in observed ILI (we expect that spikes in ILI activity on holidays are partially due to the provider composition in ILINet changes for that week -- more clinics are closed and thus ED providers have a relatively higher contribution).

Modeling across geographic scales in a single, unified model ensures forecasts are simultaneously coherent, a feature that is not present in many FluSight submissions. Ongoing work by our team will provide a model-agnostic tool by which users can modify outputs from a non-unified model so as to attain coherency. Our team is also working to incorporate internet data sources (i.e., nowcasting) into future iterations of Dante. When internet data sources were incorporated into DBM the performance increased, which leads us to be hopeful that Dante will also be improved by the thoughtful incorporation of internet data.

\section*{Acknowledgements}
This work was funded by the Department of Energy at Los Alamos National Laboratory under contract 89233218CNA000001 through the Laboratory-Directed Research and Development Program, specifically LANL LDRD grant 20190546ECR. The authors thank C.C. Essix for her encouragement and support of this work, as well as the helpful conversations with the Delphi group at Carnegie Mellon University. Approved for unlimited release under LA-UR-19-28977.



\beginsupplement

\title{Supplemental Information: \\ Multiscale Influenza Forecasting}
\date{}


\maketitle

\normalsize

\section{Dante's State-level Model}\label{sec_main:Dante}
Dante's state-level model is a Bayesian, hierarchical model composed of hierarchically specified random walks and reverse random walks.\blfootnote{$^*$Email correspondence can be directed to dosthus@lanl.gov}
Dante's state-level model is decomposed into a process model and a data model specified conditionally on the process model. In what follows, we describe all the modeling details of Dante's state-level model.

Throughout Dante's specification, pragmatic prior distributional and hyperparameter choices were made that we believe are, on balance, reasonable.
We feel justified in this position, given Dante's winning performance in the CDC's prospective 2018/19 FluSight challenge.
That said, it is possible further improvements to Dante's forecasting performance could be achieved through a more rigorous investigation into choices of priors and hyperparameters.
A rigorous investigation could include performing formal cross-validation over combinations of prior distributions and hyperparameters.

\subsection{Data Model}\label{sec:data_model}
\begin{figure}[tbhp]
	\centering
	\includegraphics[width=.95\linewidth]{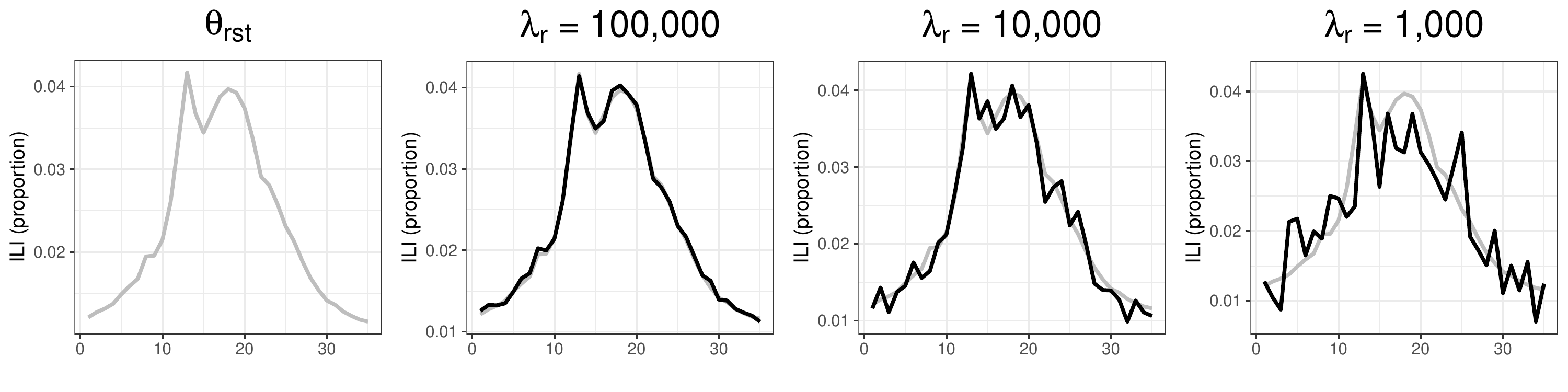}
	\caption{The underlying mean of Dante's data model (left) and three realizations from Dante's data model under three different values of $\lambda_r$. When $\lambda_r$ is large, realizations (black line) closely match the underlying mean process (grey line). As $\lambda_r$ gets smaller, realizations from the data model deviate more significantly from the underlying mean process.}
	\label{fig:lambda}
\end{figure}

Let $y_{rst} \in (0,1)$ represent observed influenza-like illness (ILI) as a proportion for state $r=1,2,\ldots,R$, for flu season $s=1,2,\ldots,S$, and week of season $t=1,2,\ldots,T$. Dante models ILI as

\begin{align}
y_{rst}|\theta_{rst}, \lambda_r &\sim \text{Beta}(\lambda_r \theta_{rst}, \lambda_r(1-\theta_{rst})),
\end{align}
\noindent where $\theta_{rst} \in (0,1)$. That is, observed ILI/100 is modeled with a Beta distribution where

\begin{align}
\text{E}(y_{rst}|\theta_{rst}, \lambda_r ) &= \theta_{rst},\\ 
\text{Var}(y_{rst}|\theta_{rst}, \lambda_r) &= \frac{\theta_{rst}(1-\theta_{rst})}{1 + \lambda_r}.
\end{align}
\noindent The conditional variance of $y_{rst}$ is modeled with a state-specific parameter $\lambda_r > 0$, allowing different states to have different variances. 
The data $y_{rst}$ is a proportion, where the denominator of the proportion is the number of patient visits reported by ILINet providers for any reason. 
Different states have widely different volumes of patient visits reported by ILINet, ranging from less than 1000 a week by the US Virgin Islands to just under 100,000 a week in Virginia and New York City. 
The noise in those estimates is related to those patient visits, with noisier estimates associated with smaller patient visit volumes. 
The noise in the estimated proportions are captured by the parameter $\lambda_r$, as illustrated in Figure \ref{fig:lambda}. 
As $\lambda_r$ decreases, the variance of the Beta distribution increases, resulting in noisier realizations from the data model.

The parameter $\lambda_r,$ i.e. the state-specific variance parameter, is modeled hierarchically

\begin{align}
\lambda_r | \lambda_{\text{prec}} &\sim \text{t}_{[0,\infty)}(0, \lambda_{\text{prec}}, 3), \\
\lambda_{\text{prec}} &\sim \text{Gamma}(5,5),
\end{align}

\noindent where $\text{t}_{[0,\infty)}$ is a half, non-standardized t-distribution with support in the interval $[0,\infty)$. 
The degrees of freedom parameter is set to be 3 to ensure that the mean and variance for the non-standardized t-distribution are defined while still maintaining a heavy-tailed distribution. 
The non-centrality parameter is set to 0, corresponding to a central, non-standardized t-distribution.
The Gamma(5,5) prior for $\lambda_{\text{prec}}$ was selected so that the prior expectation for $\lambda_{\text{prec}}$ is 1.
A non-standardized t-distribution with non-centrality parameter equal to 0 and $\lambda=1$ corresponds to the Student's t-distribution with only a degrees of freedom parameter.
The parameter $\lambda_r$ was not modeled explicitly as a function of number of patient visits as 1) the form of the relationship between $\lambda_r$ and patient visit volume was not known \emph{a priori} and 2) the number of future patient visits are not available for forecasting purposes.


\subsection{Process Model}\label{sec:process_model}
The process model is a model for the quantity $\theta_{rst}$, the latent expected ILI proportion for state $r$ in season $s$ for week of season $t$. Dante model's $\theta_{rst}$ as

\begin{align}
\theta_{rst} &= \text{logit}^{-1}(\pi_{rst}),\\
\pi_{rst} &= \mu^{\text{all}}_t + \mu^{\text{state}}_{rt} + \mu^{\text{season}}_{st} + \mu^{\text{interaction}}_{rst}.
\end{align}
\noindent That is, Dante models $\theta_{rst}$ as a function of the sum of four components: $\mu^{\text{all}}_t$, $\mu^{\text{state}}_{rt}$, $\mu^{\text{season}}_{st}$, and $\mu^{\text{interaction}}_{rst}$. The models for those components are subsequently described.

\subsubsection{Process Model: $\mu^{\text{all}}_t$}
The term $\mu^{\text{all}}_t$ is the anchor of the process model. It captures structure across the flu season common to all states and seasons. Specifically, $\mu^{\text{all}}_t$ is modeled as a random walk. For $t=1$,

\begin{align}
\mu^{\text{all}}_1 | \sigma^{2,\text{common}}_1  & \sim \text{N}(0, \sigma^{2,\text{common}}_1), \\
\sigma^{2,\text{common}}_1 | \tau^{\text{common}}_1 &\sim \text{N}_{[0,\infty)}(0,1/\tau^{\text{common}}_1),\\
\tau^{\text{common}}_1 &\sim \text{Gamma}(5,5),
\end{align}
\noindent where $\text{N}_{[0,\infty)}$ is a half Normal distribution. For $t=2,3,\ldots,T$,

\begin{align}
\mu^{\text{all}}_t | \mu^{\text{all}}_{t-1}, \sigma^{2,\text{common}} &\sim \text{N}(\mu^{\text{all}}_{t-1}, \sigma^{2,\text{common}}),\\
\sigma^{2,\text{common}} | \tau^{\text{common}}&\sim \text{N}_{[0,\infty)}(0,1/\tau^{\text{common}}),\\
\tau^{\text{common}} &\sim \text{Gamma}(5,5).
\end{align}




\subsubsection{Process Model: $\mu^{\text{state}}_{rt}$}

The term $\mu^{\text{state}}_{rt}$ captures the state-specific deviation from $\mu^{\text{all}}_{t}$, capturing effects common to all seasons within a state but distinct across states. The term $\mu^{\text{state}}_{rt}$ is modeled as a hierarchical random walk.

For $r=1,2,\ldots,R$, $\mu^{\text{state}}_{r,1}$ is modeled as

\begin{align}
\mu^{\text{state}}_{r,1} | \sigma^{2,\text{state}}_0 &\sim \text{N}(0,\sigma^{2,\text{state}}_0),\\
\sigma^{2,\text{state}}_0 | \tau^{\text{state}}_0 &\sim \text{N}_{[0,\infty)}(0, 1/\tau^{\text{state}}_0), \\
\tau^{\text{state}}_0 &\sim \text{Gamma}(5,5).
\end{align}

\noindent For $r=1,2,\ldots,R$ and $t=2,3,\ldots,T$,

\begin{align}
\mu^{\text{state}}_{rt} | \mu^{\text{state}}_{r,t-1}, \sigma^{2,\text{state}}_r &\sim \text{N}(\mu^{\text{state}}_{r,t-1},\sigma^{2,\text{state}}_r).
\end{align}

\noindent For $r=1,2,\ldots,R$,

\begin{align}
\sigma^{2,\text{state}}_r  | \lambda^{\text{state}}&\sim \text{t}_{[0,\infty)}(0, \lambda^{\text{state}},3),
\end{align}
\noindent with 

\begin{align}
\lambda^{\text{state}} &\sim \text{Gamma}(5,5).
\end{align}





\subsubsection{Process Model: $\mu^{\text{season}}_{st}$}
The term $\mu^{\text{season}}_{st}$ captures the season-specific deviation from $\mu^{\text{all}}_{t}$, capturing effects common to all states within a season but distinct across seasons. The term $\mu^{\text{season}}_{st}$ is modeled as a reverse random walk. The reverse random walk requires a prior specification for the final week of the season, $T$, rather than the first week of the season as is required with a random walk. Terms indexed by season $s$ are prone to large uncertainty intervals when used for forecasting, as no data are available to constrain their trajectories. By modeling $\mu^{\text{season}}_{st}$ as a reverse random walk and placing a prior on time $T$, we effectively turn our forecasting extrapolation problem into an interpolation problem. Though the flu season is highly variable in the middle of the season, it is quite well-constrained at the end of the season (e.g., the end of May). This makes specifying a prior at the end of the flu season a safer proposition than it may at first appear.

For $s=1,2,\ldots,S$, the model for $\mu^{\text{season}}_{sT}$ is

\begin{align}
\mu^{\text{season}}_{sT} | \sigma^{2, \text{season}}_T &\sim \text{N}(0,\sigma^{2, \text{season}}_T),\\
\sigma^{2, \text{season}}_T |  \lambda^{\text{season}} &\sim \text{t}_{[0,\infty)}(0, \lambda^{\text{season}}, 3),\\
\lambda^{\text{season}} &\sim \text{Gamma}(5,5).
\end{align}

\noindent For $s=1,2,\ldots,S$ and for $t=1,2,\ldots,T-1$,

\begin{align}
\mu^{\text{season}}_{st} | \mu^{\text{season}}_{s,t+1}, \sigma^{2, \text{season}} &\sim \text{N}(\mu^{\text{season}}_{s,t+1},\sigma^{2, \text{season}}),\\
\sigma^{2, \text{season}} | \sigma^{2, \text{season}}_T, \lambda^{\text{season}} & \sim \text{t}_{[0,\sigma^{2, \text{season}}_T]}(0, \lambda^{\text{season}}, 3).
\end{align}
\noindent The model for $\sigma^{2,\text{season}}$ is constrained to be less than $\sigma^{2,\text{season}}_T$ helping constrain the trajectories of $\mu^{\text{season}}_{st}$ from wandering wildly.




\subsubsection{Process Model: $\mu^{\text{interaction}}_{rst}$}  
The term $\mu^{\text{interaction}}_{rst}$ captures the season-specific deviation from $\mu^{\text{all}}_{t} + \mu^{\text{state}}_{rt} + \mu^{\text{season}}_{st}$, capturing structure that is specific to season $s$ and state $r$. Dante models $\mu^{\text{interaction}}_{rst}$ as a hierarchical reverse random walk. 

For $r=1,2,\ldots,R$ and $s=1,2,\ldots,S$, $\mu^{\text{interaction}}_{rsT}$ is modeled as

\begin{align}
\mu^{\text{interaction}}_{rsT} | \eta^{\text{interaction}}_{r}, \sigma^{2,\text{interaction}}_{rT}  &\sim \text{N}(\eta^{\text{interaction}}_{r}, \sigma^{2,\text{interaction}}_{rT}), \label{eq:mu_interaction} \\
\eta^{\text{interaction}}_{r} | \sigma^{2,\text{interaction}}_0 &\sim \text{N}(0, \sigma^{2,\text{interaction}}_0), \\
\sigma^{2,\text{interaction}}_0 &\sim \text{N}_{[0,\infty)}(0,20),
\end{align}
\noindent where the variance of 20 was selected as a wide, uninformative prior for $ \sigma^{2,\text{interaction}}_0$. For $r=1,2,\ldots,R$,  $s=1,2,\ldots,S$ and $t=1,2,\ldots,T-1$,   

\begin{align}
\mu^{\text{interaction}}_{rst} | \mu^{\text{interaction}}_{rs,t+1}, \alpha^{\text{interaction}}_r, \sigma^{2,\text{interaction}}_{rt} &\sim \text{N}(\alpha^{\text{interaction}}_r \mu^{\text{interaction}}_{rs,t+1},  \sigma^{2,\text{interaction}}_{rt}). \label{eq:mu_revwalk}
\end{align}

\noindent The autoregressive term $\alpha_r^{\text{interaction}} \in (0,1)$ helps regularize $\mu_{rst}^{\text{interaction}}$ towards 0. 
Because $\mu_{rst}^{\text{interaction}}$ is modeled as a reverse-random walk, as we step backward in time from $T$ to $T-1$ all the way to current time $t$ (i.e., $T$ minus some $k \geq 0$), the shrinking effect on 

$$\text{E}(\mu^{\text{interaction}}_{rst}|  \mu^{\text{interaction}}_{rs,t+1}, \alpha^{\text{interaction}}_r, \sigma^{2,\text{interaction}}_{rt} )$$
due to $\alpha_r^{\text{interaction}} \in (0,1)$ increases with distance from $T$ (i.e., with increasing $k$). 
Specifically, by the law of iterated expectations and noting all expectations are implicitly conditioned on $\eta_r^{\text{interaction}}$, $\alpha_r^{\text{interaction}}$, and $\sigma_{rt}^{2,\text{interaction}}$ for $t=1,2,\ldots,T$, we have the following:

\begin{align*}
\mathrm{E}(\mu^{\text{interaction}}_{rsT}) &= \eta^{\text{interaction}}_{r} \tag*{(Equation (\ref{eq:mu_interaction}))}, \\
\mathrm{E}(\mu^{\text{interaction}}_{rs,T-1}) &=  \mathrm{E}(\mathrm{E}(\mu^{\text{interaction}}_{rs,T-1}|\mu^{\text{interaction}}_{rsT}) ) \\
&= \mathrm{E} (\alpha_r^{\text{interaction}} \mu^{\text{interaction}}_{rsT}) \tag*{(Equation (\ref{eq:mu_revwalk}))} \\
&= \alpha_r^{\text{interaction}} \mathrm{E} (\mu^{\text{interaction}}_{rsT}) \\
& = \alpha_r^{\text{interaction}} \eta^{\text{interaction}}_{r}, \\
&\vdots\\
\mathrm{E}(\mu^{\text{interaction}}_{rs,T-k}) &=  \mathrm{E}(\mathrm{E}(\mu^{\text{interaction}}_{rs,T-k}|\mu^{\text{interaction}}_{rs,T-(k+1)}) ) \\
&= \mathrm{E} (\alpha_r^{\text{interaction}} \mu^{\text{interaction}}_{rs,T-(k+1)}) \\
&= \alpha_r^{\text{interaction}} \mathrm{E} (\mu^{\text{interaction}}_{rs,T-(k+1)}) \\
&\vdots\\
&= (\alpha_r^{\text{interaction}})^k \mathrm{E} (\mu^{\text{interaction}}_{rsT})  \\
&= (\alpha_r^{\text{interaction}})^k  \eta^{\text{interaction}}_{r}.
\end{align*}

\noindent We found this regularization helps the forecast intervals from getting too large, as $\mu_{rst}^{\text{interaction}}$ is the most challenging term in Dante to learn as only data from the state $r$ and season $s$ directly constrain it. For forecasting purposes, we have no or only partial data available for season $s$. In general, more care needs to be taken to model and constrain components indexed by $s$, as there will be little data to constrain those terms when Dante is used for forecasting.


For $r=1,2,\ldots,R$, the model for the autoregressive term $\alpha_r^{\text{interaction}}$ is,

\begin{align}
\alpha^{\text{interaction}}_r | \nu_\text{a}^{\text{interaction}}, \nu_\text{b}^{\text{interaction}} &\sim \text{Beta}(\nu_\text{a}^{\text{interaction}}, \nu_\text{b}^{\text{interaction}}),\\
\nu_\text{a}^{\text{interaction}} &\sim \text{Gamma}(5,5),\\
\nu_\text{b}^{\text{interaction}} &\sim \text{Gamma}(5,5).
\end{align}
\noindent The prior means for $\nu_\text{a}^{\text{interaction}}$ and $\nu_\text{a}^{\text{interaction}}$ are both 1.
A Beta(1,1) is a Uniform(0,1) distribution.
Thus, the prior choices for $\alpha^{\text{interaction}}_r$ reflect a lack of prior information for the autoregressive term $\alpha^{\text{interaction}}_r$.

Finally, for $r=1,2,\ldots,R$ and $t=1,2,\ldots,T$,

\begin{align}
\sigma^{2,\text{interaction}}_{rt} |  \lambda^{\text{interaction}} &\sim \text{t}_{[0,\infty)}(0, \lambda^{\text{interaction}}, 3),\\
\lambda^{\text{interaction}} &\sim \text{Gamma}(5,5).
\end{align}






\section{MCMC and JAGS}\label{sec:mcmc}
Samples from the posterior distributions and posterior predictive distributions of Dante are drawn via Markov chain Monte Carlo (MCMC). We use the software JAGS (Just Another Gibbs Sampler) as called from the \texttt{rjags} package in \texttt{R}. We run three MCMC chains, each run for 30,000 iterations, keeping every 10th iteration. We throw out the first 1,500 thinned samples as burnin, resulting in 1,500 samples per chain, or a total of $M = $ 4,500 samples per JAGS run. 

Below is the JAGS code used to fit and sample from Dante. The following quantities are passed into JAGS:
\begin{itemize}
	\item \verb NT ~is a scalar indicating how many weeks are to be modeled in the flu season. It is set equal to 35.
	\item \verb NR_state ~is a scalar for the number of states to be modeled. It is set equal to 54, including Florida.
	\item \verb NS ~is a scalar for the number of flu seasons to be modeled. It is set equal to 8, corresponding to the 8 seasons inclusively between the 2010/11 and 2017/18 seasons.
	\item \verb nobs ~is a scalar denoting how many weeks of the flu season have been observed at the time of forecasting. It is set to an integer between 5 and 29, inclusively.
	\item \verb fcstseason ~is a scalar denoting the index for the flu season being forecasted. It is set to an integer between 3 and 8, inclusively. Due to missing data, we do not forecast seasons 2010/11 and 2011/12. However, the available data for those two seasons are used for fitting.
	\item \verb yobs_nat ~is a vector of length \texttt{nobs} with the national wILI/100 estimates corresponding to the first \texttt{nobs} weeks of the \texttt{fcstseason} flu season.
	\item \verb yobs_region ~is an \texttt{nobs} $\times$ \texttt{10} matrix, with each row corresponding to a week of the flu season and each column corresponding to an HHS Region. The \texttt{yobs\_region[t,r]} entry is the wILI/100 estimate corresponding to week \texttt{t} of flu season \texttt{fcstseason} for HHS Region \texttt{r}.
	\item \verb yobs_state ~is an \texttt{NR\_state} $\times$ \texttt{NS}  $\times$ \texttt{nobs} dimensional array where the first dimension indexes states, the second dimension indexes flu season, and the third dimension indexes week of flu season. The \texttt{yobs\_state[r,fcstseason,t]} entry is the ILI/100 estimate for state \texttt{r} of flu season \texttt{fcstseason} of week of flu season \texttt{t}.
	\item \verb y_state ~is an \texttt{NR\_state} $\times$ \texttt{NS}  $\times$ \texttt{NT} dimensional array where the first dimension indexes states, the second dimension indexes flu season, and the third dimension indexes week of flu season. The \texttt{y\_state[r,s,t]} entry is the ILI/100 estimate for state \texttt{r} of flu season \texttt{s} of week of flu season \texttt{t}. If the ILI/100 estimate is missing for index \texttt{[r,s,t]}, the missing value indicator \texttt{NA} is used. When ILI/100 is very close to zero, JAGS struggles with numerical stability issues. Thus, for all non-missing ILI/100 values, we set all ILI/100 less than 0.0005 to 0.0005.
	\item \verb census_weights ~is an \texttt{NR\_state} $\times$ \texttt{11} matrix where each column sums to 1 and each entry is greater than or equal to 0. The first 10 columns correspond to the 10 HHS Regions. The 11th column corresponds to the nation. Each row corresponds to a state. The \texttt{census\_weights[r,i]} entry is the relative weight state \texttt{r} has in geographic region \texttt{i}, as determined by 2010 US Census population estimates. If the entry \texttt{census\_weights[r,i]} equals 0, that means state \texttt{r} is not in geographic region \texttt{i}.
\end{itemize}

The JAGS code for Dante, state-level model and aggregation model, is below.

\footnotesize
\begin{verbatim}
model{

#############################################
## AGGREGATION MODEL

#############################################
## national forecast model
for(t in (nobs+1):NT){
futurey_nat[t] <- t(census_weights[,11]) %*% futurey_state[,t]
}
for(t in 1:nobs){
futurey_nat[t] <- yobs_nat[t]
}

#############################################
## region forecast model
for(r in 1:10){
for(t in (nobs+1):NT){
futurey_hhs[r,t] <- t(census_weights[,r]) %*% futurey_state[,t]
}
}
for(r in 1:10){
for(t in 1:nobs){
futurey_hhs[r,t] <- yobs_region[t,r]
}
}

#############################################
## draw from state-level posterior predictive distribution
for(r in 1:NR_state){
for(t in (nobs+1):NT){
futurey_state[r,t] ~ dbeta(lambda[r] * pi_all[r,fcstseason,t], 
lambda[r] * (1 - pi_all[r,fcstseason,t]))
}
}
for(r in 1:NR_state){
for(t in 1:nobs){
futurey_state[r,t] <- yobs_state[r,fcstseason,t]
}
}

##############################################
## DANTE'S DATA MODEL

##############################################
## state data model
for(r in 1:NR_state){
for(s in 1:NS){
for(t in 1:NT){
## data model
y_state[r,s,t] ~ dbeta(lambda[r] * pi_all[r,s,t], 
lambda[r] * (1 - pi_all[r,s,t]))

## data mean
pi_all[r,s,t] <- ilogit(mu_all[t] + mu_season[s,NT+1-t] + mu_state[r,t] + mu_interaction[r,s,NT+1-t])
}
}
}  

for(r in 1:NR_state){
lambda[r] ~ dt(0, lambda_prec, 3) T(0,)
}
lambda_prec ~ dgamma(5, 5)

###########################################
## DANTE'S PROCESS MODEL

###########################################
## mu_all time series
mu_all[1] ~ dnorm(0, pow(var_all_init, -1))

for(t in 2:NT){
mu_all[t] ~ dnorm(mu_all[t-1], pow(var_all,-1))
}

var_all_init  ~ dnorm(0, prec_all_init) T(0,)
var_all       ~ dnorm(0, prec_all) T(0,)
prec_all_init ~ dgamma(5, 5)
prec_all      ~ dgamma(5, 5)

###########################################
## mu_season time series

for(s in 1:NS){
mu_season[s,1] ~ dnorm(0, pow(var_season_init, -1))
}

for(s in 1:NS){
for(t in 2:NT){
mu_season[s,t] ~ dnorm(mu_season[s,t-1], pow(var_season, -1))
}
}

var_season_init   ~ dt(0, prec_season, 3) T(0,)
var_season        ~ dt(0, prec_season, 3) T(0, var_season_init)
prec_season       ~ dgamma(5, 5)

###########################################
## mu_state time series

for(r in 1:NR_state){
mu_state[r,1] ~ dnorm(0, pow(var_state_init, -1))
}

for(r in 1:NR_state){
for(t in 2:NT){
mu_state[r,t] ~ dnorm(mu_state[r,t-1], pow(var_state[r], -1))
}
}
for(r in 1:NR_state){
var_state[r] ~ dt(0, prec_state, 3) T(0,)
}
prec_state ~ dgamma(5, 5)

var_state_init  ~ dnorm(0, prec_state_init) T(0,) 
prec_state_init ~ dgamma(5, 5)

###########################################
## mu_interaction time series

## mu_interaction initialization
for(r in 1:NR_state){
for(s in 1:NS){
mu_interaction[r,s,1] ~ dnorm(mu_interaction_mean[r], pow(var_interaction[r,1], -1))
}
mu_interaction_mean[r] ~ dnorm(0,pow(var_interaction_mean,-1))
}
var_interaction_mean ~ dnorm(0,pow(.05,-1)) T(0,)

for(r in 1:NR_state){
for(s in 1:NS){
for(t in 2:NT){
mu_interaction[r,s,t]  ~  dnorm(alpha_interaction[r] * mu_interaction[r,s,t-1], pow(var_interaction[r,t], -1)) 
}
}
}
for(r in 1:NR_state){
alpha_interaction[r] ~ dbeta(alpha_interaction_a, alpha_interaction_b)
}
alpha_interaction_a ~ dgamma(5, 5)
alpha_interaction_b ~ dgamma(5, 5)

## prior over interaction variances
for(r in 1:NR_state){
for(t in 1:NT){
var_interaction[r,t]  ~ dt(0, prec_interaction, 3) T(0,)
}
}
prec_interaction ~ dgamma(5, 5)
}
\end{verbatim}

\normalsize
\section{Details for Computing Average Standardized Week-to-Week (w)ILI Volatility}
\label{sec:standardized_volatility}

Average standardized week-to-week (w)ILI volatility for region $r$, $v_r$, displayed in Figure \ref{fig:eda_scale}, is computed as follows:

\begin{align}
v_r &= S^{-1} \sum_{s=1}^{S} v_{rs},
\end{align}
\noindent where

\begin{align}
v_{rs}&= \sqrt{(T-1)^{-1} \sum_{t=2}^T (y^*_{rst}- y^*_{rs,t-1})^2}.
\end{align}
\noindent Furthermore,

\begin{align}
y^*_{rst} &= (y_{rst} - \bar{y}_{rs})/ \sigma_{rs},
\end{align}
\noindent where

\begin{align}
\bar{y}_{rs} &= T^{-1}\sum_{t=1}^T y_{rst},\\
\sigma_{rs} &= (T-1)^{-1}\sum_{t=1}^T (y_{rst} - \bar{y}_{rs})^2,
\end{align}
\noindent $S$ is the number of flu seasons, and $T$ is the number of epidemic weeks. 
$y^*_{rst}$ is the region/season standardized (w)ILI which has a mean of zero and a standard deviation of one by construction. 
Without standardizing, regions with higher (w)ILI would appear to be more volatile simply because they have higher levels of (w)ILI. 
The average standardized week-to-week volatility for state $r$ represents a measure of how volatile the ILI time series is for that state.

\section{State ILI}

\begin{sidewaysfigure*}
	\centering
	\includegraphics[width=1\linewidth]{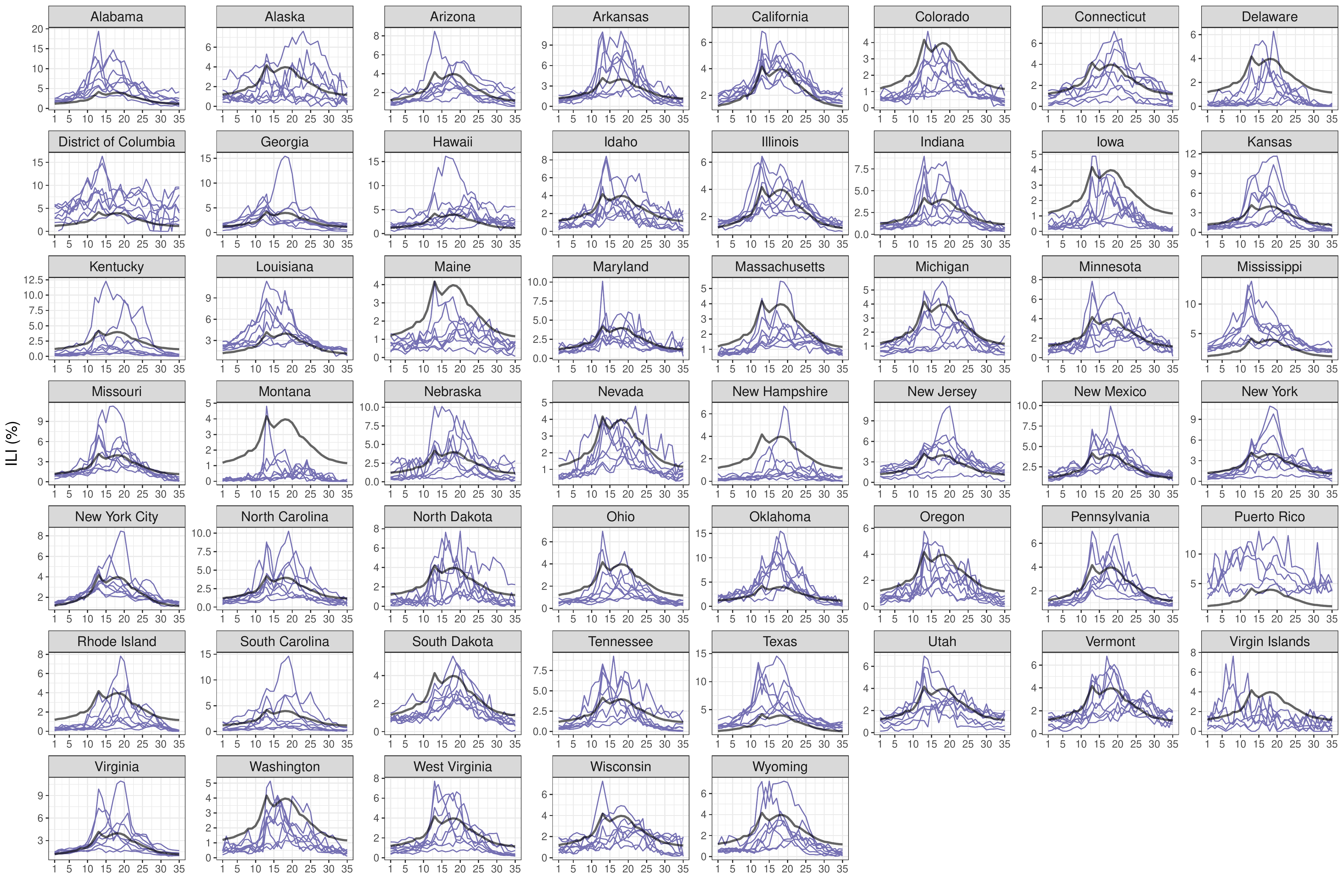}
	\caption{ILI for the 2010 through 2017 flu seasons (grey lines) for all states. Black line is the average wILI nationally for the same time period. The weeks of the flu season (x-axis) run from roughly the first week of October (x=1) through the end of May (x=35). }
	\label{fig:eda_all_states}
\end{sidewaysfigure*}

Figure \ref{fig:eda_all_states} displays ILI for all states for the 2010 through the 2017 flu seasons. The black line represents the average national wILI over that same time period for reference. Significant state-to-state variability exists, with some states having ILI consistently above the national average (e.g., Alabama, Oklahoma, Puerto Rico, Texas), and some states consistently below the national average (e.g., Montana, New Hampshire, Oregon, Rhode Island). Some states are more volatile (e.g., North Dakota, Puerto Rico) while some are less volatile (e.g., Virginia, New York City).

\section{Census weighted ILI to wILI}

\begin{sidewaysfigure*}
	\centering
	\includegraphics[width=1\linewidth]{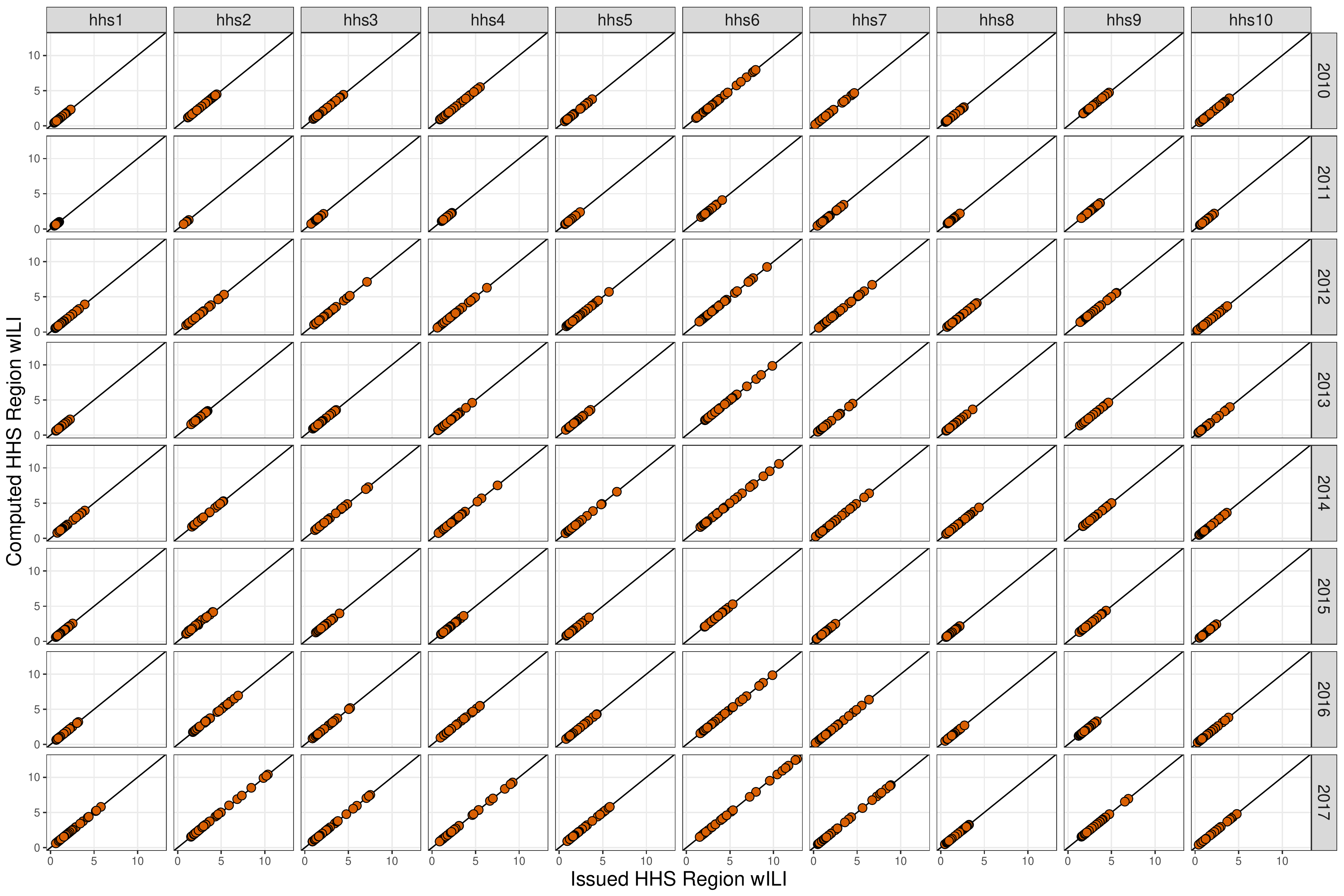}
	\caption{Health and Human Services (HHS) Region weighted influenza-like illness (wILI) issued by the Centers for Disease Control and Prevention (CDC) on the x-axis against HHS Region wILI computed as the weighted combination of state-level ILI issued by the CDC  on the y-axis. The weights are proportional to the 2010 US Census population estimates. Computed and issued HHS Regional wILI are nearly identical. Each panel corresponds to an HHS Region (column) and a flu season (row).}
	\label{fig:aggregation}
\end{sidewaysfigure*}

Information about HHS Regions is available at \url{www.hhs.gov/about/agencies/iea/regional-offices/index.html}. Figure \ref{fig:aggregation} shows wILI as reported by the CDC on the x-axis against the 2010 US Census weighted constructed wILI, calculated as a weighted average of state-level ILI with weights proportional to 2010 US Census population counts on the y-axis. Very close agreement is shown, indicating that wILI can be computed as a weighted combination of state-level ILI.

\subsection{Census weighted ILI to wILI example}

For concreteness, we show that wILI on EW49 of 2017 for HHS Region 9, reported as \textbf{2.596}, can be computed as the weighted combination of state ILI. HHS Region 9 has four states: Arizona, California, Hawaii, and Nevada. The 2010 US Census population estimates for the four HHS Region 9 states are 6,407,774 (AZ), 37,320,903 (CA), 1,363,963 (HI), and 2,702,464 (NV), for a total of 47,795,104 people in HHS Region 9. The state weights are then 0.134 (AZ), 0.781 (CA), 0.029 (HI), and 0.057 (NV). CDC reported ILI for EW49 of 2017 for those states are 3.284 (AZ), 2.498 (CA), 4.341 (HI), and 1.434 (NV).

The state-weighted wILI estimate for HHS Region 9 is:

\begin{align*}
\text{w}_{\text{AZ}}*\text{ILI}_{\text{AZ}} +  \text{w}_{\text{CA}}*\text{ILI}_{\text{CA}} +  \text{w}_{\text{HI}}*\text{ILI}_{\text{HI}} +  \text{w}_{\text{NV}}*\text{ILI}_{\text{NV}}&= \\
0.134 * 3.284 + 0.781 * 2.498 + 0.029 * 4.341 + 0.057 + 1.434 &= \bm{2.596},
\end{align*}
\noindent the same wILI value reported by the CDC.

\section{Forecasting targets}

The forecasting targets as selected by the CDC for the FluSight challenge are defined as follows:
\begin{itemize}
	\item Short-term forecasts ($n-$week ahead):
	\begin{itemize}
		\item The (w)ILI percentage for each target week $n$ weeks ahead of the most recently released (w)ILI data, where $n=1,\ldots,4$. 
	\end{itemize}
	\item Peak intensity:
	\begin{itemize}
		\item The highest value of (w)ILI in a given season.
	\end{itemize}
	\item Peak timing: 
	\begin{itemize}
		\item The week(s) in which the highest value of (w)ILI in a given season is reached.
	\end{itemize}
	\item Onset:
	\begin{itemize}
		\item The first week in which (w)ILI reaches or exceeds the given regional/national baseline value for three consecutive weeks in a season. Note that `baseline,' a threshold used to define epidemic onset, is calculated by adding two standard deviations to the mean observed ILI levels in non-flu weeks during the previous three influenza seasons; see \url{https://www.cdc.gov/flu/weekly/overview.htm}.
	\end{itemize}
\end{itemize}

Due to backfill, the CDC sets a validation date at which to record all of the above `validation' values for each forecast challenge. For this paper, data for all seasons were collected from the CDC’s website on Friday, October 12th of 2018 and validation \%ILI values are considered those present in the collected data set.

The ILINet national and regional baseline values are available at \url{www.cdc.gov/flu/weekly/overview.htm}. Note that state data do not have defined historic baseline values.

\begin{figure*}
	\centering
	\includegraphics[width=1\linewidth]{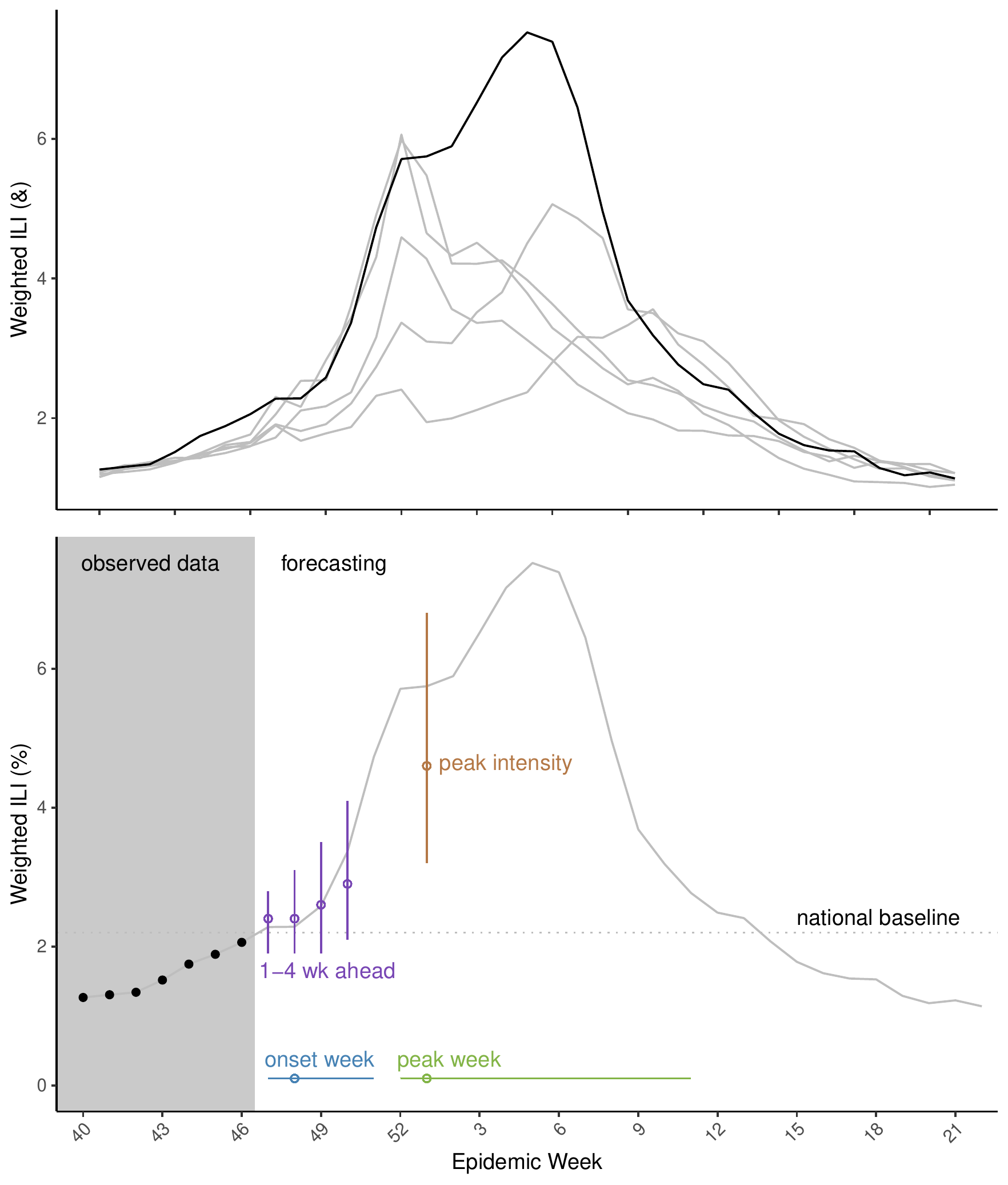}
	\caption{Top: national wILI for the 2012 through 2017 flu seasons, i.e. those used in the leave-one-season-out training and testing of DBM and Dante. The 2017 flu season, the hold-out season considered in the lower plot, is emphasized in black. Bottom: Dante's national forecasts for the CDC targets in 2017 made with (w)ILI data from epidemic week 46 being the last available data. Posterior mode and 95\% HPD interval for each target are shown as an open point and line, respectively. For these forecasts, the model is utilizing all data from the 2012 through 2016 flu seasons, and the first 7 weeks of data from the 2017 flu season (indicated by the solid black points). The remainder of the 2017 flu season (shown in the bottom figure as a grey line) has yet to be observed.}
	\label{fig:targets_plot}
\end{figure*}

A visual example of Dante's national-level predictions for each of the forecasting targets may be found in Figure \ref{fig:targets_plot}.

\section{Scoring procedure}
\label{sec_main:Scoring_proc}

Scoring protocols used in this paper correspond to the stated scoring procedure used by the CDC's FluSight challenge for the 2018/19 season. Complete national and regional FluSight challenge description and scoring guidelines may be found by navigating to the `Guidance Documents' page from \url{predict.cdc.gov/post/5ba1504e5619f003acb7e18f}. State FluSight guidelines are available by navigating to the `Guidance Documents' page from \url{https://predict.cdc.gov/post/5ba5389fa983f303b832726b}.

All forecasts are evaluated using the validation \%ILI data. That is, the larger effects of backfill should have subsided and the ILINet observations used are stable from week-to-week. This lack of backfill leads to a relatively higher performance of both Dante and DBM relative to their performance in a real-time FluSight challenge.

\subsection{Scoring by target}

Note that for the purposes of finding the peak timing, onset, and scoring, all (w)ILI values are rounded to the nearest tenth. For example, a reported (w)ILI of 5.387\% would become 5.4\%. The scoring bin width is one tenth of a percentage for short-term targets and peak intensity, and one week for peak timing and season onset.

\begin{itemize}
	\item Short-term forecasts ($n-$week ahead):
	\begin{itemize}
		\item The probability assigned to the correct bin plus the probabilities assigned to the five preceding and proceeding bins are summed.
		\item For example, if (w)ILI at week 49 is 2.5\%, the probabilities assigned to all bins ranging from 2.0\% to 3.0\% inclusively are summed in the 1-week-ahead forecast from week 48 to get that week's 1-week ahead forecast skill.
	\end{itemize}
	\item Peak intensity:
	\begin{itemize}
		\item The probability assigned to the correct bin plus the probabilities assigned to the five preceding and proceeding bins are summed.
		\item For example, if (w)ILI peaks at 5.4\%, the probabilities assigned to all bins ranging from 4.9\% to 5.9\% inclusively are summed to get the skill for peak timing.
	\end{itemize}
	\item Peak timing: 
	\begin{itemize}
		\item The probability assigned to the correct week(s) plus the probabilities assigned to the immediately preceding and proceeding weeks are summed.
		\item In the case of multiple peaks, the probability assigned to each correct week plus the probability assigned to the non-overlapping preceding and proceeding weeks for each peak is summed; each week's probability is only counted once.
		\item For example, if (w)ILI peaks at 5.4\% on weeks 3 and 5, then the probabilities assigned to weeks 2, 3, 4, 5, and 6 are summed to get the skill for peak timing.
	\end{itemize}
	\item Onset:
	\begin{itemize}
		\item The probability assigned to the correct week(s) plus the probability assigned to the immediately preceding and proceeding weeks is summed.
		\item For example, if the onset is week 48, then the probabilities assigned to weeks 47, 48, and 49 are summed to get the skill for onset.
		\item Note that onset is not defined at the state level.
	\end{itemize}
\end{itemize}

For all targets, if the correct week/bin is near the first or last possible bin then the total number of weeks/bins summed for scoring is reduced accordingly. For example, if the validation (w)ILI value is 0.3\%, then only bins ranging from 0\% to 0.8\% will be counted for scoring $n-$week ahead predictions (there won't be extra bins counted above 0.8\% to `make up for' the decrease in the number of lower bins).

\subsection{Averaging scores}

In this paper, forecasts are made beginning at epi-time 5 (i.e., epi-week 45, after four weeks of data from a given season are available) and the last forecast is made at epi-time 29 (i.e., after the influenza season has long been over). 

When calculating an overall score for a given location, season, and/or model, we take the geometric mean of a subset of the scores for that location/season/model. That is, relevant log skills across the targets of interest are averaged and the result is exponentiated. However, the weeks at which a forecast is made and the target chosen impact whether or not a given week's forecast skill is included. The scoring inclusion window used in this paper is the intersection of this paper's forecast window (epi-time 5 through 29) and the CDC definition for weeks to score (discussed below).
Note that when skill is 0 (i.e., when the probability assigned to the scorable bin(s) by the model is 0), the log score is set to -10 for that target.

\subsubsection{CDC's Evaluation Period}

For state forecasts, all weeks are included in the average score calculation because there are no state-level baseline values. For regional and national forecasts, the CDC chooses evaluation periods based on utility of said forecasts. For all seasonal targets (i.e. onset, peak timing, and peak intensity), the CDC evaluation period begins with the first forecast submission. For onset, the CDC evaluation period ends six weeks after the validation onset. For peak timing and intensity, the evaluation period ends after (w)ILI is observed to go below baseline for the final time during an influenza season (we interpret this to mean the scored weeks are those up to and including the first week below baseline after which all weeks are below baseline). For short-term forecasts (i.e., $n-$week ahead predictions) the CDC evaluation period begins four weeks prior to the observed onset week and ends three weeks after (w)ILI is observed to go below baseline for the final time during an influenza season. We assume all windows are inclusive (i.e., three weeks after implies that the third week out \textit{is} counted). The idea is that forecasts within these windows are those that would be most practically useful.

\subsubsection{Averaging scores example}

As an example, suppose the goal is to generate an overall score for Dante's onset forecasts for HHS6 in 2014. The historical baseline for HHS6 in 2014 was 3.2, and the observed onset week was week 47 (epi-time 8). The CDC evaluation period would include forecasts made during epi-time 1 through 14, inclusive. We produce model forecasts during epi-time 5 through 29. The overall score would be the geometric mean (i.e., the exponentiated log average score) for HHS6 onset forecasts made on epi-time 5 through 14, inclusive, in 2014. For the purposes of illustration, we round the following values. Specifically, the skill at epi-time 5 (week 44) is 0.27, meaning that in week 44 the sum of the probability assigned to weeks 46, 47, and 48 (i.e., the validation onset week and its preceding and proceeding week) was 0.27. At epi-time 6, the skill is 0.22. At epi-times 7 through 14 the skills are 0.10, 0.68, 0.99, 0.99, 0.99,  0.99, 0.99, and 0.99. Thus the summary score for onset of the 2014/2015 season of flu in HHS6 is:

$$e^{\frac{\text{ln}(0.27) + \text{ln}(0.22) + \text{ln}(0.10) + \text{ln}(0.68) + \text{ln}(0.99) + \text{ln}(0.99) + \text{ln}(0.99) + \text{ln}(0.99) + \text{ln}(0.99) + \text{ln}(0.99)}{10}} = 0.57$$

In order to get an overall score for Dante's forecasts for HHS6 across all targets and seasons, the geometric mean would be taken of all of the relevant scores from each season-target combination.

\section{Data cleaning}

The state data were pulled from the CDC’s website on Friday, October 12th of 2018 before being cleaned. In the data, the \code{ili} variable is the percent of patients having ILI symptoms out of the total number of patients seen. For the majority of observations, the \code{ili} variable is equivalent to 100 times the \code{ilitotal} variable divided by the \code{total\_patients} variable. When \code{total\_patients}$=0,$ \code{ili} is set to \code{NA} because the true proportion of ILI among the population is unknown. However, there are cases in which there are \code{ili = NA} estimates, even though the \code{total\_patients} $> 0$ and \code{ilitotal}$ = 0$. We change the value of \code{ili} in this scenario (i.e., the scenario when \code{total\_patients} $> 0$ and \code{ilitotal}$ = 0$) to be 0 rather than \code{NA} as there is information about the ILI estimate. It just happens to be an estimate of 0.

\begin{table}[hptb!]
	\centering
	\begin{tabular}{lllllllllll}
		\code{ili} & \code{ilitotal} & \code{total\_patients} &  &  &  &  &  & \code{ili} & \code{ilitotal} & \code{total\_patients} \\
		0.586      & 3               & 512                    &  &  &  &  &  & 0.586      & 3               & 512                    \\
		0.242      & 1               & 413                    &  &  &  &  &  & 0.242      & 1               & 413                    \\
		\code{NA}  & 0               & 373                    &  &  &  &  &  & 0.000      & 0               & 373                    \\
		\code{NA}  & 0               & 0                      &  &  &  &  &  & \code{NA}  & 0               & 0                     
	\end{tabular}
	\caption{(Left) The relationship between \code{ili}, \code{ilitotal}, and \code{total\_patients} pre-cleaning. Note that in the third row \code{ili} is \code{NA} but \code{total\_patients}$>0$. (Right) The relationship between \code{ili}, \code{ilitotal}, and \code{total\_patients} post-cleaning. In the third row the \code{NA} has been replaced with a 0. The \code{NA} in the fourth row persists because for this observation \code{total\_patients} is 0.}
	\label{tab:datacleaning}
\end{table}

Note that in recent releases of (w)ILI data, the presence/absence of \code{NA} coding may differ from that used in older data releases. Should this model be used for new data, cleaning said data with the logic used above for when (w)ILI should be 0 vs. \code{NA} is recommended.

\subsection{Missing state-level data}

\begin{figure*}
	\centering
	\includegraphics[width=1\linewidth]{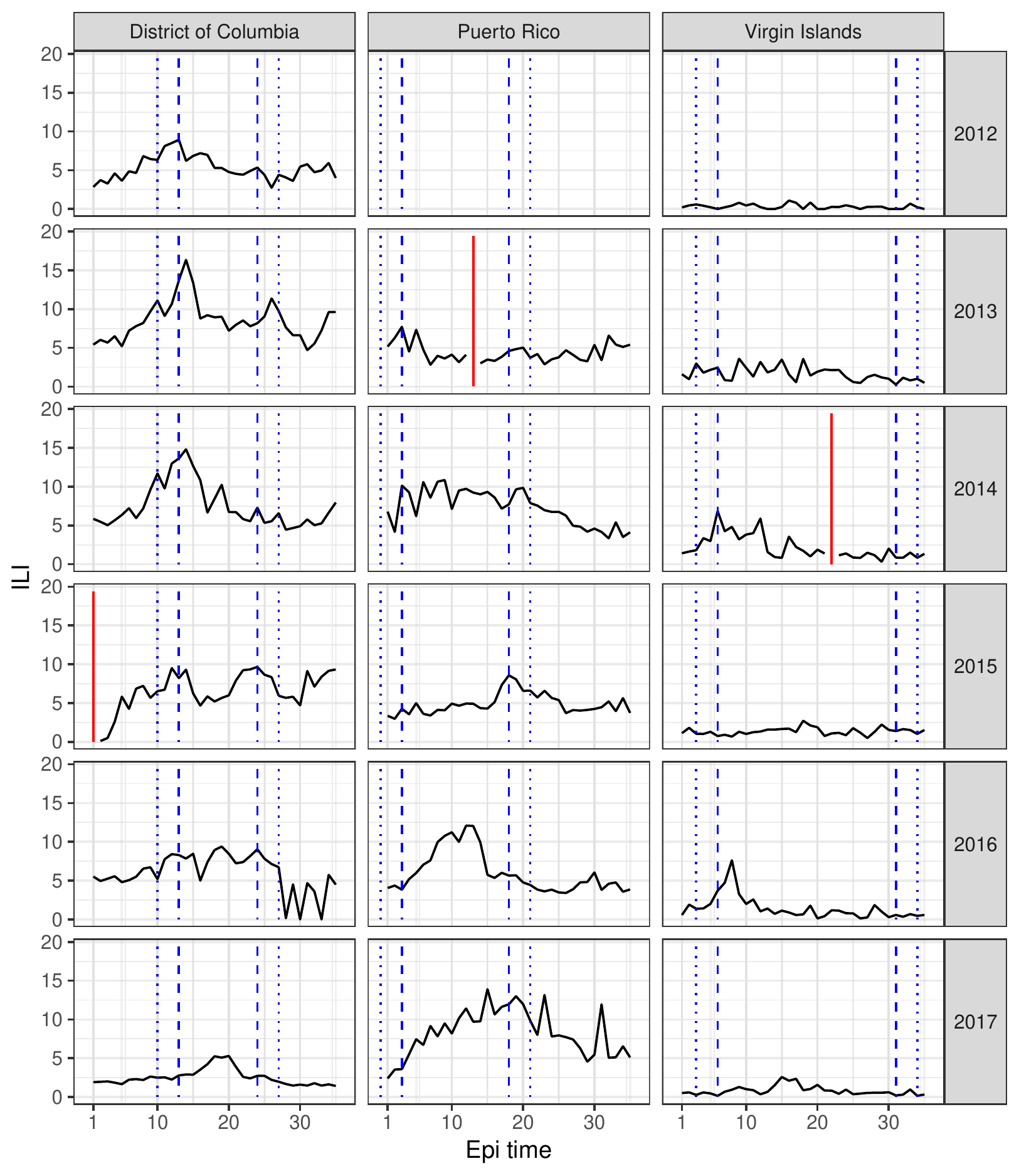}
	\caption{The three missing observations among state-season combinations which are not completely unobserved are denoted by the red vertical lines in each subplot. The inner long dashed blue lines show the minimum and maximum historic peak weeks from the 2010/2011 to 2017/2018 seasons, while the outer short dashed blue lines show the three week buffer about this historic window. The validation ILI values are shown in black. For Puerto Rico in 2013 and the Virgin Islands in 2014, the \code{NA} occurs within the buffered historic window and leads the seasonal targets in those state-seasons to be un-scorable. In the District of Columbia in 2015, on the other hand, the \code{NA} occurs outside the buffered historic window and so seasonal targets for that state-season are scorable.}
	\label{fig:miss_states}
\end{figure*}

After using the above data cleaning techniques, 3 missing state-season-week observations persisted (see Figure \ref{fig:miss_states}), excluding completely missing state-seasons (e.g., Florida, for which no data were available for any state-season-week, and Puerto Rico in 2012). For a given state-season, if any weeks are missing data it is theoretically possible for that week to be the validation peak week. Thus, the seasonal targets of peak timing and peak intensity have \code{NA} for a ground truth in instances in which that state-season has any missing data and the forecast skill for those seasonal targets must also be \code{NA}.

Practically speaking, however, not every missing week need preclude scoring seasonal targets for a whole state-season. For example, it is incredibly unlikely for influenza to peak in mid-May. On the other hand, a missing week in mid-December right near the highest observed ILI value has a much higher chance of being the validation peak. For determining which weeks should be considered `irrelevant' missing weeks, we calculate the historic minimum and maximum observed peak timing from the 2010/2011 season to the 2017/2018 season for each state. We then utilize this state-specific min/max historic peak timing minus/plus a three week `buffer' as the (inclusive) window in which an \code{NA} will lead to that state-season seasonal forecasts are determined to be un-scorable.

\subsection{Missing region-level data}

There are no \code{NA}s present in the region-level data. When a state-season-week is missing for a given region, that state is omitted from the weighted average calculation for that region's wILI and the weights are adjusted accordingly.

\subsection{Model output cleaning}
As mentioned previously, a skill of 0 means that the log-score is set to -10 for that target per CDC guidelines. In practice, we place a small probability in every bin, to eliminate the possibility of incredibly damaging -10s. For peak week and onset, we choose 0.00018 to be the smallest probability in every bin, where the worst possible log score will then be $\text{ln}(3*.00018) \approx -7.5$ (using the multibin log score); for peak percentage and short-term forecasts, we choose .00005 so the worst possible log score will be about $\text{ln}(11*.00005) \approx -7.5.$ Note that these particular padding values were chosen as a simple risk mitigation step. Further study could be done to determine some ``optimal’’ pad for scoring, but for the purposes of this paper the main goal was comparability of Dante and DBM so the same numeric pad was used for each.

\section{Forecast skill}

We consider detailed results for Danta's forecast skill for each prediction time broken down by target-season at the national level and the state level for two example states. Discussion is based on Figures \ref{fig:skill_national}--\ref{fig:example_wILI}. Note that in Figures \ref{fig:skill_national}--\ref{fig:skill_montana} the x-axis shows the time at which the forecast was made, while the (w)ILI data that are overlaid have been shifted to the left $n$ weeks for each $n$-week ahead skill plot. For example, for the 1-week-ahead forecast skill plots the skill plotted at epi time 5 is the skill in forecasting (w)ILI at epi time 6, while the scaled inverse (w)ILI plotted at epi time 5 is the corresponding (w)ILI value from epi time 6.

\subsection{National skill}

\begin{figure*}
	\centering
	\includegraphics[width=1\linewidth]{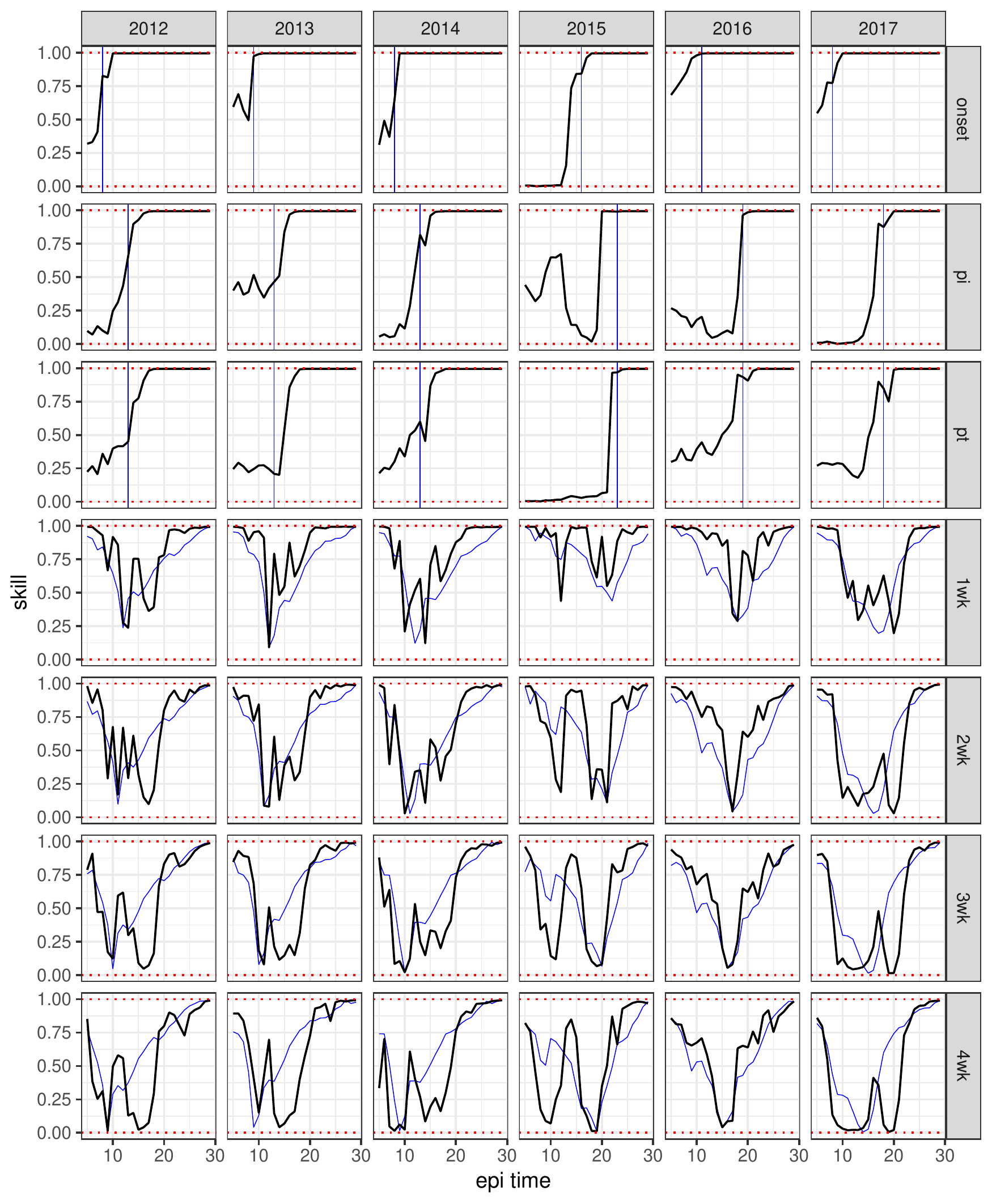}
	\caption{Dante's forecast skill by prediction time for each target at the national level (black line). The validation onset (peak timing) for each season is shown in the first (second/third) row as a vertical blue line. Shifted national wILI data for each season multiplied by $-1$ is shown as the thin blue line overlaid on the short-term target subplots (scaled so as to have the same range as skill in each subplot) For $n-$week ahead forecasts $-$wILI is shifted back $n$ weeks such that $-$wILI for the week being predicted is overlaid on the week in which the prediction was made. It is evident that short-term target skill is inversely related to wILI.}
	\label{fig:skill_national}
\end{figure*}

Figure \ref{fig:skill_national} shows Dante's forecast skill for each prediction time broken down by target-season. Early in the season, the seasonal targets each tend to have low scores; by the end of the season (after the seasonal metric has been observed) the scores increase to near 1. The short-term forecasts are much more variable, but broadly skill is inversely related to wILI.

Consider the first row of Figure \ref{fig:skill_national}. By at most three weeks after an onset has been observed (the defining feature of onset being the consistent wILI above baseline for three weeks), the onset skill reaches near 1 because the model uses the observed wILI data as its prediction upon observation and there is no backfill to contend with, unlike with current-season forecasts. Of note is that for 2015, which was an unusually late-peaking season, the onset skill is near 0 early in the season because at this point the model is basing its predictions on previous seasons' trajectories, meaning that a low-flu late-peaking season has a low probability of occurring before observing much data from the current season. 

The final four rows of Figure \ref{fig:skill_national} illustrate the inverse relationship between skill and wILI. This general phenomenon is likely partly due to the nature of the scoring for short-term targets. Recall that the probabilities assigned to the bins from -0.5 to 0.5 away from the validation wILI level are summed. If the validation wILI level is very high, the model has more room to be `wrong' even if it is generally right that levels are high. For example, if the validation wILI one week ahead is 7 and the model puts most of its weight around a wILI of 6, the model will score very poorly even though the concept of high wILI is correct. On the other hand, if the validation wILI is low and the model predicts that wILI is low, more probability will tend to be near the truth simply because there are fewer `low' wILI values. The higher skill during early/late season is likely also partly due to the fact that wILI behaves much more consistently at the beginning/end of seasons when flu is not circulating as heavily, and week-to-week changes tend to be smaller and occur more gradually, so the trajectory learned from previous seasons and other states in the model is more useful to a new given state-season combination at the tails of the season. 

\subsection{State skill}

\begin{figure*}
	\centering
	\includegraphics[width=1\linewidth]{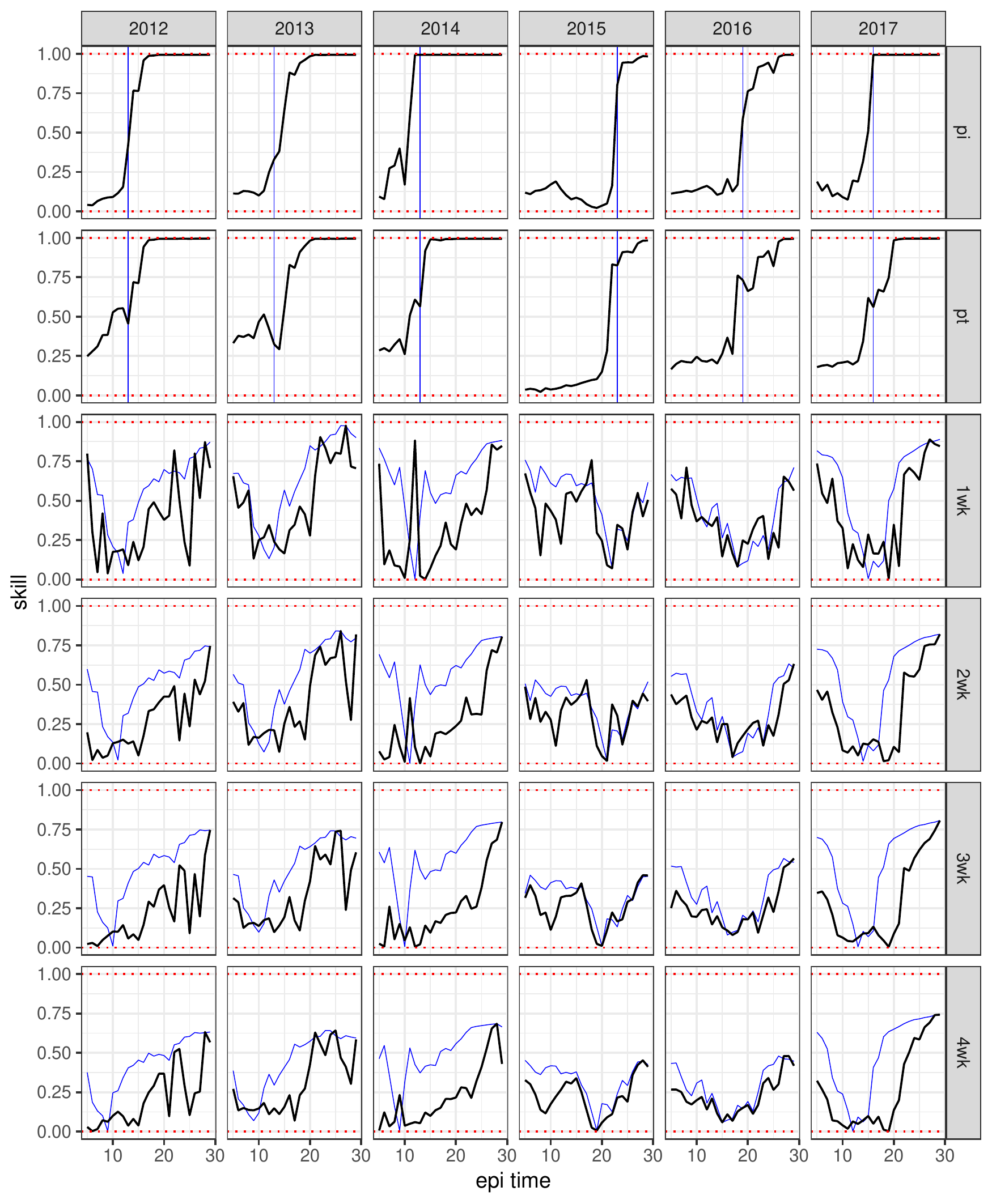}
	\caption{Dante's forecast skill by prediction time for each target in Alabama (black line). The validation onset (peak timing) for each season is shown in the first (second/third) row as a vertical blue line. Shifted ILI data for each season multiplied by $-1$ is shown as the thin blue line overlaid on the short-term target subplots (scaled so as to have the same range as skill in each subplot) For $n-$week ahead forecasts $-$ILI is shifted back $n$ weeks such that $-$ILI for the week being predicted is overlaid on the week in which the prediction was made. Short-term target skill tends to decrease as ILI increases.}
	\label{fig:skill_alabama}
\end{figure*}

\begin{figure*}
	\centering
	\includegraphics[width=1\linewidth]{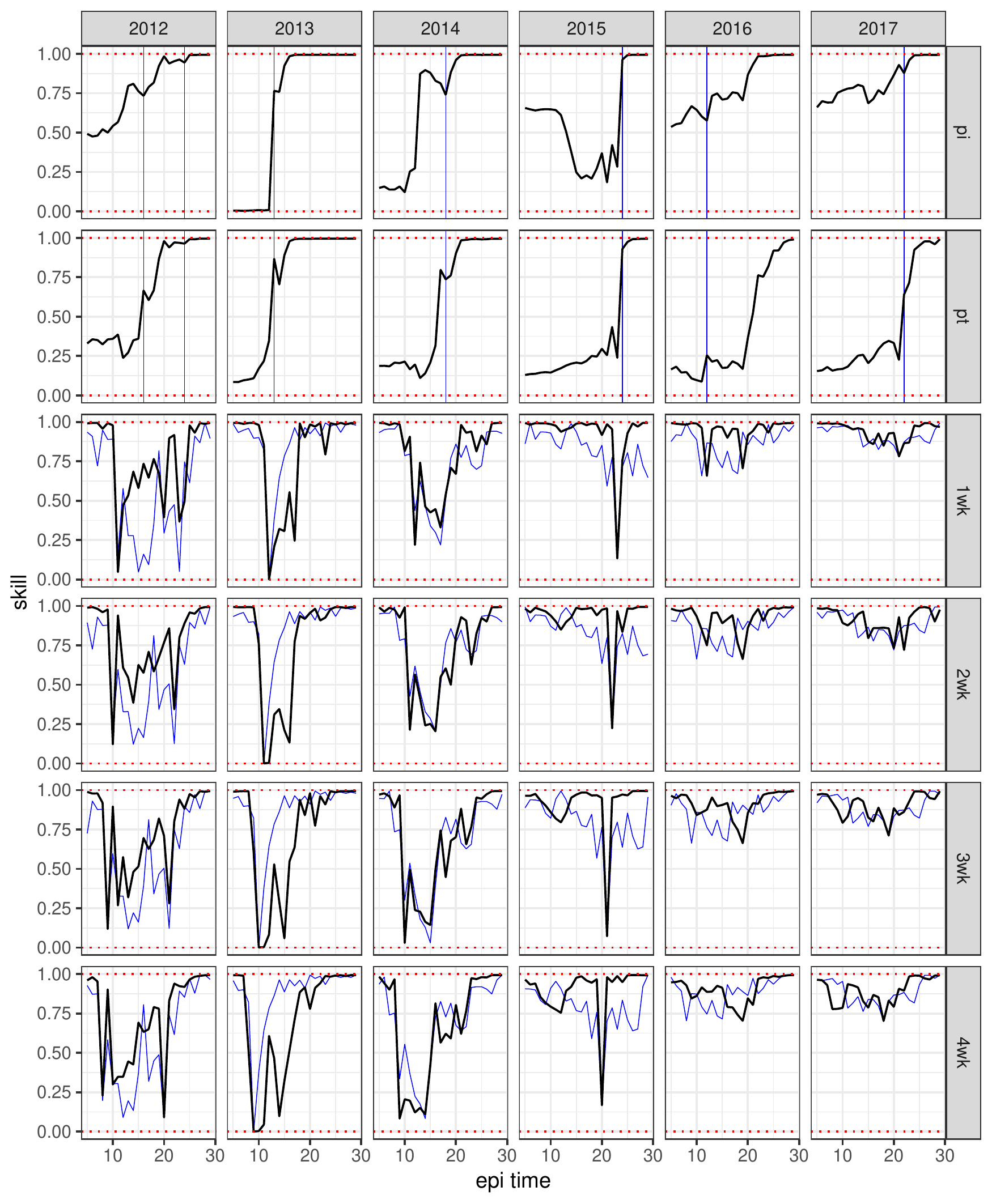}
	\caption{Dante's forecast skill by prediction time for each target in Montana (black line). The validation onset (peak timing) for each season is shown in the first (second/third) row as a vertical blue line or lines. Shifted ILI data for each season multiplied by $-1$ is shown as the thin blue line overlaid on the short-term target subplots (scaled so as to have the same range as skill in each subplot) For $n-$week ahead forecasts $-$ILI is shifted back $n$ weeks such that $-$ILI for the week being predicted is overlaid on the week in which the prediction was made.}
	\label{fig:skill_montana}
\end{figure*}

\begin{sidewaysfigure*}
	\centering
	\includegraphics[width=1\linewidth]{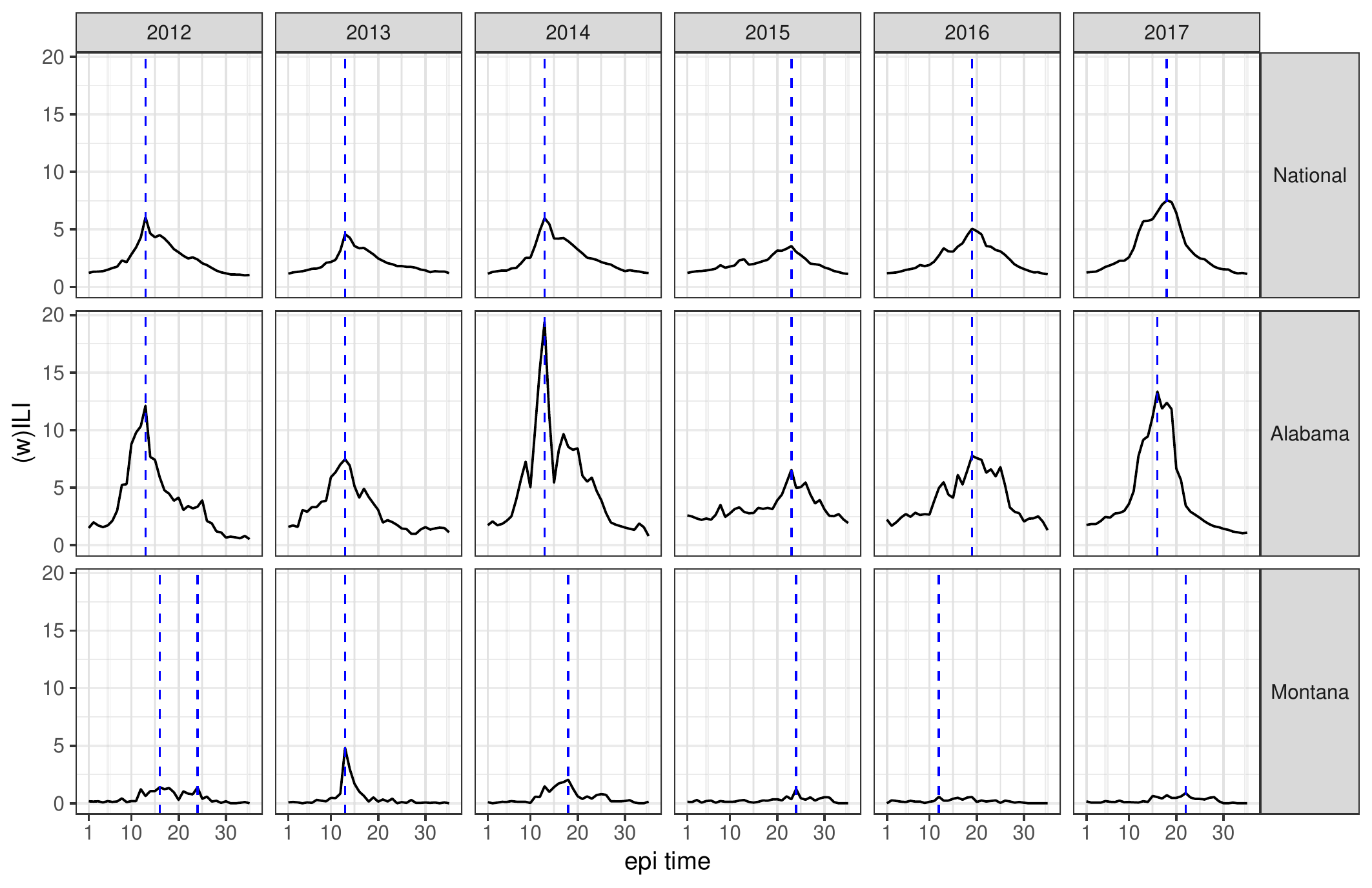}
	\caption{National wILI data and state ILI data for the example skill plots discussed. This figure is intended as a reference for magnitude of (w)ILI data because the -(w)ILI data in the skill plots has been scaled to lie in the same range as skill. Note that Alabama has exceptionally high ILI relative to the national average, while Montana has particularly low ILI relative to the national average. The vertical dashed blue line marks the peak week(s) for each region-season combination.}
	\label{fig:example_wILI}
\end{sidewaysfigure*}

Figures \ref{fig:skill_alabama} and \ref{fig:skill_montana} show target-season skill plots for two example states (Alabama and Montana, respectively). The same general trends with seasonal target skill for national data are found in the state-level seasonal target skills with some notable differences. For many target-season combinations, the rise of state-level peak incidence/timing (pi/pt) skill to near 1 occurs later after the peak has occurred than for the national pi/pt skill. For example consider the pi/pt skill for Montana in the 2016 season (the first and second rows in the second from the rightmost column in Figure \ref{fig:skill_montana}). After the peak has occurred, it is nearly the end of the season before pi/pt skill rises to near 1. Looking at the bottom row in the second from the rightmost column in Figure \ref{fig:example_wILI}, which shows the validation ILI for Montana in 2016, it is evident that this was a particularly early/weak-peaking and low-magnitude season for Montana. The model had trouble telling that a peak had in fact occurred, so some probability remained allocated to later peak timing and higher peak incidence. In Montana, pi/pt skill rises almost immediately as the peak occurs in 2015, likely because this is a fairly prominent (by Montana standards) peak occurring late in the season, so the model very readily places high probability on it being the peak after its occurrence.

The skill for short-term targets is in general much higher for Montana than for Alabama, particularly for seasons in which ILI levels in Montana lack much of a distinct peak (i.e., 2016 and 2017). The inverse relationship between ILI and and short-term forecast skill is less consistent at the state level than it was at the national level. In Alabama (the bottom four rows of Figure \ref{fig:skill_alabama}), the short-term forecasts in 2015 and 2016 track extremely well with inverse ILI; in particular, note the spiked pattern in both skill and inverse ILI from epidemic time (ET) 11 to 15 for 4-week ahead forecasts in 2016 (shifted right by one with each decrease in week, so 12 to 16 for 3-week ahead, etc.). On the other hand, the 1-, 2-, and 4-week ahead forecasts in 2014 corresponding to the peak week (made at ET 12, 11, and 9, respectively) have high skill and correspond to high ILI. When looking at the corresponding subplot in Figure \ref{fig:example_wILI}, we see that the slope to the peak is relatively constant save for the 3rd week before the peak for which the 3-week ahead forecast was poor, suggesting that the high performance may be due to the consistency of the observed path at the time of forecast with the future path. Another interesting feature of forecasts in Alabama is that short-term skill tends not to begin as high as that of national forecasts. Of note, consider the near-0 starting skill at ET 5 in 2012 for 3- and 4-week forecasts. Looking at the middle row of the leftmost column of Figure \ref{fig:example_wILI}, we see that in 2012 ILI in Alabama experienced a dip at ET 4, followed by a small rise in ET 5 and increasingly steeper rises over the subsequent 4 weeks. The 3/4-week forecast from ET 5 is unable to anticipate this rise, hence the extremely poor skill. 

\section{Accuracy comparison }
\label{sec:MSE}

\begin{figure}
	\centering
	\includegraphics[height=.9\textheight]{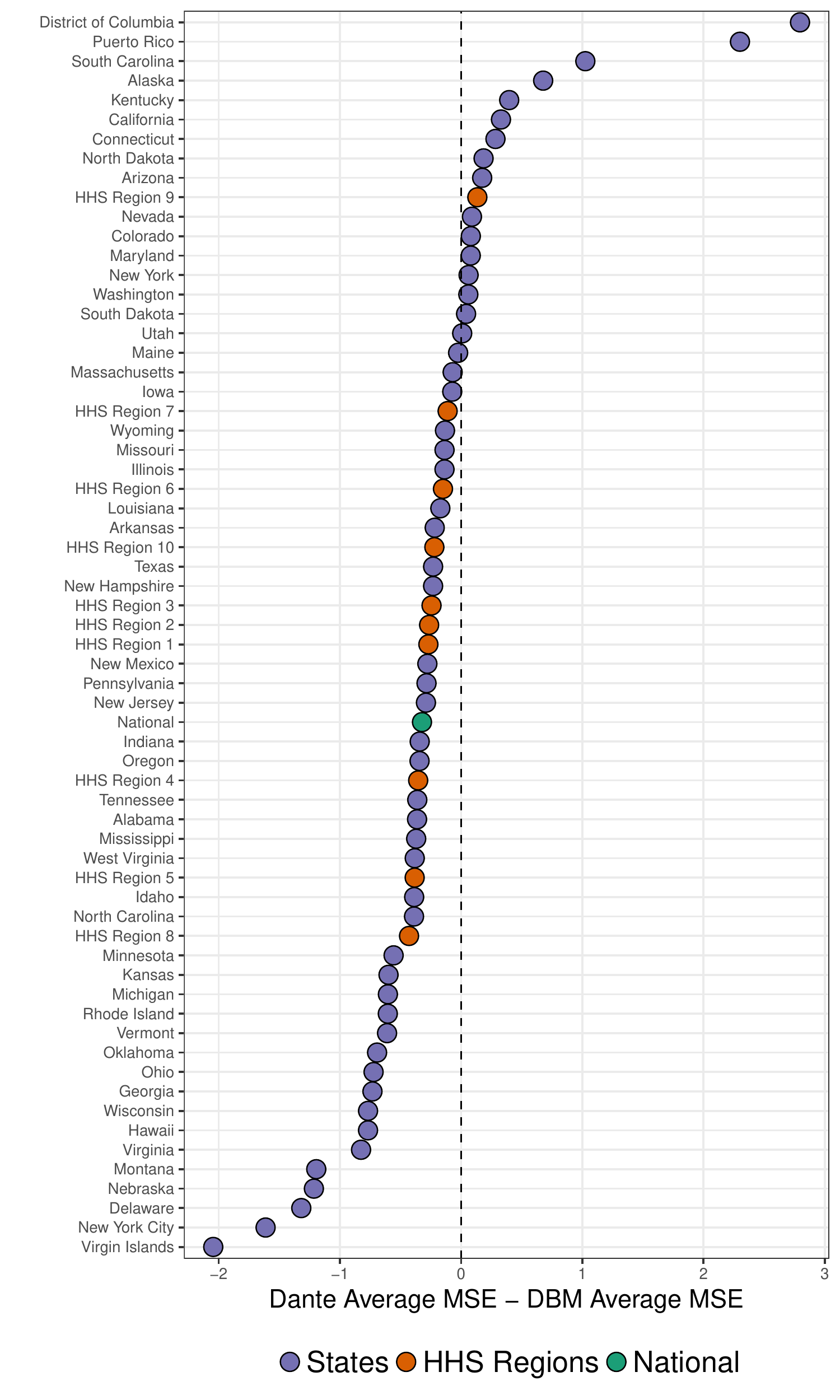}
	\caption{Difference in mean squared error (MSE) between Dante and DBM, for all states, regions, and nationally. Dante had lower MSE for the majority of geographic regions.}
	\label{fig:mse_regions}
\end{figure}

\begin{figure}
	\centering
	\includegraphics[height=.8\textheight]{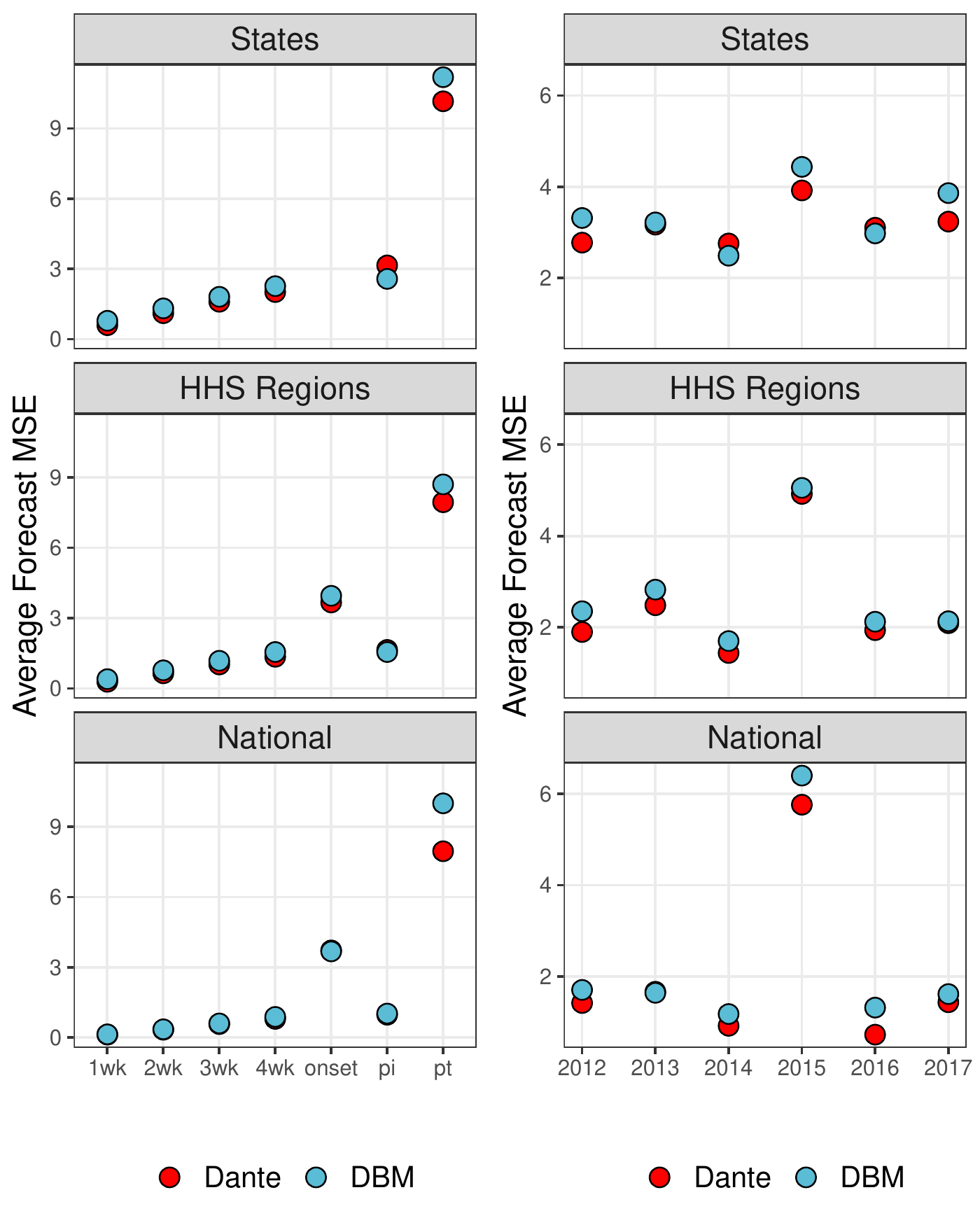}
	\caption{(left) Average mean squared error (MSE) by scales and targets. PI and PT stand for peak intensity and peak timing, respectively. Dante outperformed DBM for all scales and targets, except for onset nationally and for peak intensity (PI) regionally and by state. (right) Average forecast skill by scales and flu seasons. Dante outperformed DBM for the majority of scales and flu seasons, except for 2013 nationally and for 2014 and 2016 by state.}
	\label{fig:mse_targets_and_seasons}
\end{figure}

Accuracy is measured by mean squared error (MSE) of model point predictions.
In the main paper, we showed that Dante was both more skillful (i.e., higher forecast skill scores) and more confident (i.e., narrower 90\% HPD interval widths) than DBM. 
We also showed that Dante outperformed DBM in terms of accuracy across all geographic scales.
However, we did now show the relative performance of Dante and DBM broken down by region, season, or target. 
Figures \ref{fig:mse_regions} and \ref{fig:mse_targets_and_seasons} provide this additional information.
Note that these plots are analogous to the skill plots in the main paper, but compare MSE rather than skill.
Averages were calculated using the same set of weeks as defined by the CDC guidelines for scoring forecast skill.

Specifically, Figure \ref{fig:mse_regions} shows the difference in accuracy between Dante and DBM for each state, region, and nationally. 
Dante outperformed DBM for the majority of geographic regions, but its relative performance is less consistently better as measured by accuracy than it is by skill.

Figure \ref{fig:mse_targets_and_seasons} shows accuracy broken down by targets (left) and flu seasons (right) for each geographic scale.
Dante outperformed DBM for all scales and targets, except for onset nationally and for peak intensity (PI) regionally and by state.
Relative magnitude of differences for peak timing (PT) appear higher because the MSE are in terms of the target (i.e., MSE for PT is in terms of squared weeks).
Dante also outperformed DBM for the majority of scales and flu seasons, except for 2013 nationally and for 2014 and 2016 by state.

\section{Probability Densities}

\subsection{Normal($\mu$, $\sigma^2$)}
The probability density function for the Normal distribution is

\begin{align}
f(x|\mu, \sigma^2) = \frac{1}{2 \pi \sigma^2} \text{exp}\Bigg( -\frac{(x-\mu)^2}{2 \sigma^2} \Bigg)
\end{align}
\noindent where $x \in (-\infty, \infty)$, $\mu \in (-\infty, \infty)$, and $\sigma^2 > 0$. The mean and variance for a Normal distribution are $\text{E}(X) = \mu$ and $\text{Var}(X) = \sigma^2$.

\subsection{t($\mu, \lambda$, $k$)}
The probability density function for the non-standardized t-distribution is

\begin{align}
f(x|\mu,\lambda,k) &= \frac{\Gamma(\frac{k+1}{2})}{\Gamma(\frac{k}{2})} \Bigg(\frac{\lambda}{k \pi} \Bigg)^{\frac{1}{2}} \Bigg\{ 1 + \frac{\lambda (x-\mu)^2}{k} \Bigg\}^{-\frac{(k+1)}{2}}
\end{align}
\noindent where $x \in (-\infty, \infty)$, the non-centrality parameter $\mu \in (-\infty, \infty)$, the precision $\lambda > 0$, and the degrees of freedom $k > 0$. For $k > 1$, $\text{E}(X) = \mu$. For $k>2$, $\text{Var}(X) = \frac{k}{\lambda(k-2)}$.

\subsection{Gamma($\alpha$, $\beta$)}
The probability density function for the Gamma distribution is

\begin{align}
f(x|\alpha, \beta) &= \frac{\beta^\alpha}{\Gamma(\alpha)} x^{\alpha-1} \text{exp}(-\beta x)
\end{align}
\noindent where $x \in (0,\infty)$, $\alpha > 0$ is the shape parameter and $\beta > 0$ is the rate parameter. The mean and variance for the Gamma distribution are $\text{E}(X) = \frac{\alpha}{\beta}$ and $\text{Var}(X) = \frac{\alpha}{\beta^2}$.

\subsection{Beta($\alpha$, $\beta$)}
The probability density function for the Beta distribution is 

\begin{align}
f(x|\alpha, \beta) &= \frac{\Gamma(\alpha + \beta)}{\Gamma(\alpha) \Gamma(\beta)} x^{\alpha - 1} (1-x)^{\beta - 1}
\end{align}
\noindent where $x \in [0,1]$, $\alpha > 0$, and $\beta > 0$. The mean and variance for the Beta distribution are $\text{E}(X) = \frac{\alpha}{\alpha + \beta}$ and $\text{Var}(X) = \frac{\alpha \beta}{(\alpha + \beta)^2 (\alpha + \beta + 1)}$.

\clearpage


\end{document}